\makeatletter
\@ifundefined{@parse@version@dash}{%
\def\@parse@version#1{\@parse@version@0#1}
\def\@parse@version@#1/#2/#3#4#5\@nil{%
\@parse@version@dash#1-#2-#3#4\@nil}
\def\@parse@version@dash#1-#2-#3#4#5\@nil{%
  \if\relax#2\relax\else#1\fi#2#3#4 }
}{}
\makeatother
\documentclass[twocolumn,secnumarabic,amsmath,amssymb,aps,prr,superscriptaddress,eqsecnum,graphicx,dvipdfmx,longbibliography]{revtex4-2}

\usepackage{graphicx}
\usepackage{bm}
\usepackage{color}
\usepackage{ulem}
\usepackage[dvipdfmx,
colorlinks=true,
urlcolor=blue,
citecolor=blue,
linkcolor=blue,
hyperfootnotes=false]{hyperref}

\begin{document}
\title{
Improving accuracy of tree-tensor network approach by optimization of network structure
}

\author{Toshiya Hikihara}
\affiliation{Graduate School of Science and Technology, Gunma University, Kiryu, Gunma 376-8515, Japan}
\author{Hiroshi Ueda}
\affiliation{Center for Quantum Information and Quantum Biology, Osaka University, Toyonaka 560-0043, Japan}
\affiliation{Computational Materials Science Research Team, RIKEN Center for Computational Science (R-CCS), Kobe 650-0047, Japan}
\author{Kouichi Okunishi}
\affiliation{Department of Physics, Niigata University, Niigata 950-2181, Japan}
\author{Kenji Harada}
\affiliation{Graduate School of Informatics, Kyoto University, Kyoto 606-8501, Japan}
\author{Tomotoshi Nishino}
\affiliation{Department of Physics, Graduate School of Science, Kobe University, Kobe 657-8501, Japan}

\begin{abstract}
Numerical methods based on tensor networks have been extensively explored in the research of quantum many-body systems in recent years.
It has been recognized that the ability of tensor networks to describe a quantum many-body state crucially depends on the spatial structure of the network.
In the previous work [Hikihara {\it et al.}, Phys. Rev. Res. {\bf 5}, 013031 (2023)], we proposed an algorithm based on tree tensor networks (TTNs) that automatically optimizes the structure of TTN according to the spatial profile of entanglement in the state of interest.
In this paper, we apply the algorithm to the random XY-exchange model under random magnetic fields and the Richardson model in order to analyze how the performance of the algorithm depends on the detailed updating schemes of the structural optimization.
We then find that for the random XY model, on the one hand, the algorithm achieves improved accuracy, and the stochastic algorithm, which selects the local network structure probabilistically, is notably effective.
For the Richardson model, on the other hand, the resulting numerical accuracy subtly depends on the initial TTN and the updating schemes.
In particular, the algorithm without the stochastic updating scheme certainly improves the accuracy, while the one with the stochastic updates results in poor accuracy due to the effect of randomizing the network structure at the early stage of the calculation.
These results indicate that the algorithm successfully improves the accuracy of the numerical calculations for quantum many-body states, while it is essential to appropriately choose the updating scheme as well as the initial TTN structure, depending on the systems treated.
\end{abstract}

\date{\today}

\maketitle

\section{Introduction}\label{sec:Intro}

Since the development of the density-matrix renormalization-group (DMRG) method\cite{White1992,White1993}, the tensor-network approach has made rapid progress in a variety of fields\cite{Orus2019,OkunishiNU2022,Larsson2024}.
While tensor networks have been studied in statistical mechanics for many years\cite{Baxter1968}, the introduction of quantum-information concepts has advanced the understanding of their properties significantly, and it has been recognized that tensor networks can represent the low-energy states of quantum many-body systems efficiently.
Several algorithms based on various tensor-network states, including matrix-product state (MPS)\cite{OstlundR1995,RommerO1997}, tensor-product state (TPS)\cite{NishinoHOMAG2001,GendNishino2003} or projected entangled-pair state (PEPS)\cite{VerstraeteC2004,VerstraeteWPC2006}, and the state of multiscale entanglement renormalization ansatz (MERA)\cite{Vidal2007,EvenblyV2009}, have been developed.

A tensor network is a contraction of low-rank tensors and is utilized as an approximation of a high-rank tensor such as a wave function of a quantum many-body state.
In general, the number of elements of the high-rank tensor increases exponentially with the number of quantum spins (qudits) in the system.
One can avoid this exponential growth of the Hilbert space by 
setting an upper bound on the dimension of the auxiliary bonds in the tensor network.
The drawback of introducing the upper bound on the bond dimension is the loss of accuracy due to the truncation of the Hilbert space.
How to reduce this loss is an essential issue in tensor-network approaches.

In this paper, we explore a numerical algorithm based on the tree-tensor networks (TTNs), which are tensor networks without loops.
We particularly focus on binary trees, TTNs composed of three-leg tensors, although the extension to $m$-ary trees is straightforward.
An essential property of the TTN is that cutting an arbitrary bond in the network separates the TTN into two subsystems.
In other words, those two subsystems are connected only by the single bond.
This property of TTN leads to the following guideline for reducing the loss of accuracy due to the truncation of the Hilbert space.\cite{SekiHO2021,HikiharaUOHN2023a,OkunishiUN2023}
Since every bond in a TTN independently bridges the corresponding two subsystems, each bond must be able to express the entanglement between the subsystems in order to represent the target quantum state accurately.
However, due to the upper bound of the bond dimension, which we denote by $\chi$ in the following, the entanglement entropy (EE) that a bond can carry is upper bounded to a maximum of $\ln \chi$.
If the EE associated with each bond exceeds the upper limit, the excess entanglement is omitted in the approximated TTN and results in a loss of accuracy.
Here, recall that the number of bonds in a TTN, which is $2N-3$ where $N$ is the number of quantum spins in the system, is much smaller than the number of possible bipartitions of the system, $2^{N-1}-1$.
Therefore, by using the network structure that makes the EEs on the $(2N-3)$ bonds in the TTN as small as possible, one can minimize the loss of accuracy in approximating the target quantum state by the TTN with a finite $\chi$.
We call this guideline the least-EE principle.

The least-EE principle indicates that using a TTN with the optimal structure is crucially important for improving the accuracy of the TTN approach.
However, finding the optimal TTN structure is highly nontrivial and challenging.
Several efforts have been devoted to the problem so far.
There have been attempts to optimize the order of sites within the matrix-product network (MPN), which is employed in DMRG, based on various strategies.\cite{ChanHG2002,LegezaS2003,MoritzHR2004,LegezaVPD2015,LiRYS2022}
A method to optimize the structure of TTN using the cut-off bond dimension as a minimization function has also been proposed.\cite{Larsson2019}
For systems with randomness, the tensor-network strong-disorder renormalization group (tSDRG) method, in which a TTN suitable for a random system is constructed based on the idea of the strong-disorder renormalization group,\cite{MaDH1979,DasguptaM1980} has been developed\cite{HikiharaFS1999,GoldsboroughR2014} and successfully applied.\cite{LinKao_2017,SekiHO2020,SekiHO2021}
For small systems for which an exact wavefunction is obtained, a method to extract the optimal TTN structure from the wavefunction according to the least-EE principle has also been proposed.\cite{OkunishiUN2023}

In the previous study\cite{HikiharaUOHN2023a}, we proposed an algorithm to search for the optimal TTN structure by iterating the local reconstruction of the network based on the least-EE principle.
We applied the algorithm to some toy models for which the optimal TTN structure was deduced from the perturbative renormalization analysis\cite{HikiharaUOHN2023a,HikiharaUOHN2023b}.
We then demonstrated that the algorithm succeeded in obtaining the expected optimal structure.
In this paper, we have investigated the performance of several types of TTN structural optimization calculations, which are specified by the initial TTN structure, the scheme to select the local TTN structure in the optimization process, and the bond dimension at which the structural optimization is executed.
We have performed the calculations for
the random XY-exchange model under random magnetic fields and the Richardson model.
We have then demonstrated that for the random XY model, the structural optimization algorithm can reduce the variational energy significantly.
In particular, a stochastic scheme, which selects the local TTN structure probabilistically, can achieve the lowest variational energy regardless of the initial TTN structure.
For the Richardson model, on the other hand, we have found that while the structural optimization algorithm can yield a good TTN structure with a low variational energy, the calculations are sometimes trapped by a TTN structure unsuitable for representing the lowest energy state and fail to reach the optimal TTN structure.
These results for the two models reveal that while the TTN structural optimization algorithm is basically effective in improving the accuracy of the calculation, the performance of each type of calculation depends on the model.
The results thus indicate the importance of appropriately choosing the updating scheme of the network and the initial TTN structure according to the model to be studied.

The plan of this paper is as follows.
In Sec.\ \ref{sec:algo}, we review the structural optimization algorithm and make several comments.
Section\ \ref{subsec:scheme} describes the types of optimization calculations tested in the present study and the quantities calculated.
Sections\ \ref{subsec:random} and \ref{subsec:Rch} respectively present the numerical results for the random XY-exchange model under random magnetic fields and the Richardson model.
The performances of each type of calculation for the models are discussed there.
Section\ \ref{sec:conc} is devoted to a summary and concluding remarks.
In Appendix\ \ref{app:initialTTN}, we explain how to prepare the initial TTN.
The details of the actual calculation, including the number of sweeps, are presented in Appendix\ \ref{app:Detail_Calcu}.
Some results of the calculations with a stochastic optimization scheme are presented in Appendix\ \ref{app:Stcs2}.
Appendix\ \ref{app:Exact_Rch} describes the exact calculation of the lowest energy for the Richardson model.

\section{Algorithm}\label{sec:algo}

The algorithm we study in this work is the one proposed in Ref.\ \cite{HikiharaUOHN2023a}.
Here, we briefly review the algorithm for completeness and add comments.
We note that an algorithm to optimize the site ordering within the MPN based on a similar strategy has been proposed in Ref.\ \cite{LiRYS2022}.

\begin{figure}
\begin{center}
\includegraphics[width = 80 mm]{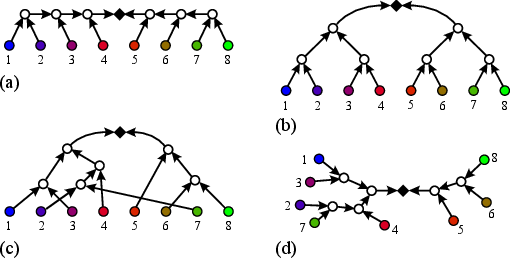}
\caption{
Examples of TTN.
(a) Matrix-Product Network (MPN),
(b) Perfect-Binary Tree (PBT),
(c) TTN of a generic form,
and (d) the same TTN as that in (c).
Open circles represent isometries and the solid diamond represents the singular-value matrix.
Circles with colors represent bare spins.
In (a), (b), and (c), the bare spins are arranged in the order of the site index.
Note that within the category of MPN and PBT, there are several networks corresponding to different site orderings.
}
\label{fig:TTNs}
\end{center}
\end{figure}

In the proposed algorithm, the wave function of the quantum state is represented by a TTN.
Figure\ \ref{fig:TTNs} shows typical examples of TTNs.
We note that the TTN includes MPN, which is employed in the DMRG method, and the perfect-binary tree (PBT).

\begin{figure}
\begin{center}
\includegraphics[width = 60 mm]{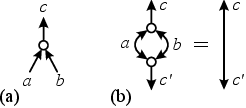}
\caption{
(a) Isometry $V_{a,b}^c$ with the legs $a$, $b$, and $c$.
(b) Schematic picture of the orthonormal condition Eq.\ (\ref{eq:orthonormal}).
Connected legs (bonds) represent a contraction of their degrees of freedom.
The double arrow on the right-hand side of (b) represents a Kronecker delta $\delta_{cc'}$.
}
\label{fig:isometry}
\end{center}
\end{figure}

Let us write a tensor in the TTN as $V_{a,b}^c$ as in Fig.\ \ref{fig:isometry} (a).
We impose the orthonormal condition on $V_{a,b}^c$,
\begin{eqnarray}
\sum_{a,b} \bar{V}_{a,b}^c V_{a,b}^{c'} = \delta_{cc'},
\label{eq:orthonormal}
\end{eqnarray}
where $\bar{V}_{a,b}^c$ is a complex conjugate of $V_{a,b}^c$,
and call the tensor the isometry.
We assign a direction to each bond in the TTN so that the bases of the bond outgoing from each tensor are orthonormal;
Namely, in Eq.\ (\ref{eq:orthonormal}), the index $c$ ($c'$) corresponds to the outgoing bond, while the indices $a$ and $b$ correspond to the incoming bonds.
All bonds in the TTN are directed from the boundary, where the bare spin degrees of freedom are, towards the ``center" of the TTN, where the diagonal matrix of singular values is located.
Such a representation format of the TTN is called the mixed canonical form\cite{ShiDV2006}.
Adopting the mixed canonical form offers strong advantages, such as guaranteeing the orthonormality of the basis of the truncated Hilbert space expanded by incoming bonds and making the contraction calculations of isometries extremely simple, as shown in Fig.\ \ref{fig:isometry} (b).
We note that the center position of the canonical form can be taken at an arbitrary bond and can also be moved by the fusing and re-decomposing process of tensors\cite{ShiDV2006}.

\begin{figure}
\begin{center}
\includegraphics[width = 73 mm]{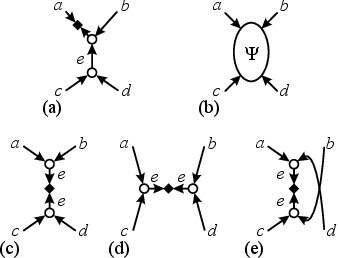}
\caption{
(a) Central area at the beginning of a step.
(b) Effective ground-state wave function $\Psi(abcd)$ obtained by the diagonalization of the effective Hamiltonian $\tilde{\mathcal{H}}$.
(c), (d), and (e) show the possible local connections of the singular-value decompositions Eqs. (\ref{eq:SVD_ab_cd}), (\ref{eq:SVD_ac_bd}), and (\ref{eq:SVD_ad_bc}), respectively.
Open circles represent isometries and a solid diamond denotes the singular-value matrix at the canonical center. 
}
\label{fig:algorithm}
\end{center}
\end{figure}

\begin{table*}
\caption{
Algorithm to optimize the structure and isometries of TTN.
}
\label{tab:algorithm}
\begin{center}
\begin{tabular}{ll}
\hline
\hline
 1.  &  Prepare an initial TTN expressed in the mixed canonical form. (See Appendix A for the detail.)\\
 2.  &  Construct the effective Hamiltonian $\tilde{\mathcal{H}}$ of the whole system on the orthonormal basis of the central area. \\
 3.  &  Diagonalize the effective Hamiltonian $\tilde{\mathcal{H}}$ to obtain the ground-state wave function $\Psi$. \\
 4.  &  Perform the singular-value decompositions of $\Psi$ for the three possible local connections in Figs.\ \ref{fig:algorithm} (c)-(e). \\
 5.  &  Choose the optimal local connection according to the adopted updating scheme of the network discribed\\
     &  in the text of Sec.\ \ref{sec:algo} and Sec.\ \ref{subsec:scheme}. \\
 6.  &  Update the local structure of TTN as well as isometries in the central area. \\
 7.  &  Move the central area according to the sweeping rule. (See Ref.\ \cite{HikiharaUOHN2023a} for the details of the rule.)\\
 8.  &  Iterate processes 2. - 7. until the sweep of the entire TTN is completed. \\
 9.  &  Iterate the sweep until the TTN structure and quantities of interest ({\it e.g.,} the ground-state energy) converge. \\
\hline
\hline
\end{tabular}
\end{center}
\end{table*}

The algorithm optimizes the structure and isometries of TTN in the following manner.
In the TTN, we focus on the ``central area", depicted in Fig.\ \ref{fig:algorithm} (a), which consists of the diagonal matrix of singular values at the canonical center located on the bond $a$ and two isometries connected to it.
The Hilbert space for the central area is expanded by the orthonormal basis of the incoming bonds $a, b, c$, and $d$, and we can construct the effective Hamiltonian $\tilde{\mathcal{H}}$ of the whole system represented in the (truncated) Hilbert space.
Then, we diagonalize the effective Hamiltonian using, {\it e.g.}, the Lanczos method to obtain the ground-state wave function $\Psi(abcd)$ [Fig.\ \ref{fig:algorithm} (b)].
Next, we perform the singular-value decomposition of $\Psi(abcd)$ for the three possible local connections of the isometries [see Figs.\ \ref{fig:algorithm} (c)-(e)],
\begin{subequations}
\label{eq:SVD}
\begin{align}
&\Psi(abcd) = \sum_{e=1}^{\chi^2} V^e_{ab} D_e V^e_{cd}, 
\label{eq:SVD_ab_cd} \\
&\Psi(abcd) = \sum_{e=1}^{\chi^2} V^e_{ac} D'_e V^e_{bd}, 
\label{eq:SVD_ac_bd} \\
&\Psi(abcd) = \sum_{e=1}^{\chi^2} V^e_{ad} D''_e V^e_{bc}.
\label{eq:SVD_ad_bc}
\end{align}
\end{subequations}
Throughout this paper, we distinguish the isometries by the letters for the bond index, if necessary.
[Hence, for example, $V^e_{ab}$ and $V^e_{cd}$ in Eq.\ (\ref{eq:SVD_ab_cd}) are different isometries.]
We also adopt the notation that the singular values are arranged in descending order, $D_1 \ge D_2 \ge \cdots \ge D_{\chi^2} \ge 0$ (the same for $\{D'_e\}$ and $\{D''_e\}$) and suppose that the ground-state wave function $\Psi(abcd)$ is normalized so that $\sum_{e=1}^{\chi^2} (D_e)^2 = \sum_{e=1}^{\chi^2} (D'_e)^2 = \sum_{e=1}^{\chi^2} (D''_e)^2 = 1$.
Since the spectra of the singular values, $\{ D_e \}$, $\{ D'_e \}$, and $\{ D''_e \}$, are different from each other, the EEs obtained from those singular values,
\begin{subequations}
\label{eq:bondEE}
\begin{align}
&\mathcal{S}^{(ab|cd)} = - \sum_{e=1}^{\chi^2} (D_e)^2 \ln (D_e)^2,
\label{eq:bondEE_ab_cd} \\
&\mathcal{S}^{(ac|bd)} = - \sum_{e=1}^{\chi^2} (D'_e)^2 \ln (D'_e)^2,
\label{eq:bondEE_ac_bd} \\
&\mathcal{S}^{(ad|bc)} = - \sum_{e=1}^{\chi^2} (D''_e)^2 \ln (D''_e)^2,
\label{eq:bondEE_ad_bc}
\end{align}
\end{subequations}
are also different.
Therefore, we can select the connection of the isometries with the smallest EE as the optimal one to update the local structure of TTN.
Simultaneously, we can update the isometries in the central area by replacing them with the tensors obtained from the adopted singular-value decomposition, where we truncate the dimension of the center bond $e$ by keeping only the bases with the $\chi$-largest singular values.\cite{Note_isometry_update}
We refer to the above procedures to update the local network structure and the isometries of the TTN [from Fig.\ \ref{fig:algorithm}(a) to Fig.\ \ref{fig:algorithm}(c)-(e)] as a ``step".
When this step is completed, we shift the central area by one isometry according to an appropriate sweeping rule of the network\cite{HikiharaUOHN2023a} and continue the computation.
We iterate the steps by sweeping the entire network to obtain the optimal TTN.
The algorithm is summarized in Table.\ \ref{tab:algorithm}.

Several comments on the algorithm are presented in order below.

In the structural optimization, the above-mentioned algorithm adopts the least-EE principle to choose the local network connection that minimizes the EE on the center bond.
There are other options here.
First, one may select the local connection that minimizes the truncated reduced density-matrix (rDM) weight\cite{LiRYS2022}, the sum of the rDM eigenvalues for the discarded bases, defined as
\begin{subequations}
\label{eq:rDMweight_cut}
\begin{align}
&\delta \rho_{\rm cut}^{(ab|cd)} = \sum_{e=\chi+1}^{\chi^2} (D_e)^2,
\label{eq:rDMweight_cut_ab_cd}\\
&\delta \rho_{\rm cut}^{(ac|bd)} = \sum_{e=\chi+1}^{\chi^2} (D'_e)^2,
\label{eq:rDMweight_cut_ac_bd}\\
&\delta \rho_{\rm cut}^{(ad|bc)} = \sum_{e=\chi+1}^{\chi^2} (D''_e)^2.
\label{eq:rDMweight_cut_ad_bc}
\end{align}
\end{subequations}
It is expected that using Eq.\ (\ref{eq:rDMweight_cut}) is advantageous to improve the computational accuracy as much as possible with a fixed $\chi$ since the truncated rDM weight is directly related to the fidelity of the variational wave function in TTN with the true ground-state wave function.
On the other hand, using the bond EE [Eq.\ (\ref{eq:bondEE})] observes the entanglement and, therefore, is expected to be suitable to grasp the entanglement geometry of the ground state.
It also bears noting that the algorithm using the truncated rDM weight should be strongly dependent on $\chi$.

Second, since the reconnection of the network performed in the algorithm is local, the structural optimization may be trapped by local minima in the landscape of network structure, especially when the system contains complexity such as randomness.
A possible strategy to avoid such traps is to implement a stochastic selection of the network structure.
Namely, one may select the local connection randomly from the three possible candidates according to the probabilities,
\begin{eqnarray}
P^{(1)} \propto \exp(-\beta \mathcal{S}),
\label{eq:stochas1}
\end{eqnarray}
where $\mathcal{S} = \mathcal{S}^{(ab|cd)}, \mathcal{S}^{(ac|bd)}, \mathcal{S}^{(ad|bc)}$ is the EE at the center bond of each local connection.
$\beta$ is an effective inverse temperature that 
is increased as the calculation proceeds.
Also, if one prefers the heavier penalty for the connection with the larger bond EE, one may use the probabilities,
\begin{eqnarray}
P^{(2)} \propto \exp(-\beta \mathcal{S}^2).
\label{eq:stochas2}
\end{eqnarray}

In the algorithm described above, one can improve only the isometries in the TTN while leaving the network structure fixed, by always choosing the local connection $(ab|cd)$ for the singular-value decomposition Eq.\ (\ref{eq:SVD_ab_cd}) at process 5 in Table\ \ref{tab:algorithm}.
Therefore, in practice, we can easily switch the calculations with and without the structural optimization of TTN.
In our calculation discussed in the subsequent sections, we first perform the calculation with structural optimization at a certain $\chi=\chi_{\rm opt}$.
Then, we continue the calculation without structural optimization by increasing $\chi$ to further improve the isometries in the TTN having the optimized structure.

In the singular-value decomposition of the wave function $\Psi(abcd)$, the number of non-zero singular values (the rank of the wave-function matrix in the decomposition considered) may be smaller than $\chi$.
This situation often occurs for the bonds close to the boundary of the TTN.
In such a case, only the bases corresponding to the non-zero singular values are adopted as the new bases for the center bond, and consequently, the dimension of the bond is smaller than the upper bound $\chi$.
In the practical calculation, we discard the bases with singular values smaller than a certain threshold (set to be $10^{-6}$ in our calculation) since their contribution to the wave function is negligibly small.
Furthermore, all the bases whose singular values are degenerate are kept or discarded altogether to preserve symmetries of the target state.
These measures may also result in an auxiliary bond with a dimension smaller than $\chi$.

In the calculation, one must be careful about the case where the dimensions of all three legs of an isometry become unity, which traps the calculation.
Suppose that the dimensions of the bonds $a$, $b$, and $e$ are all unity in the situation shown in Fig.\ \ref{fig:algorithm} (a).
Then, the singular values of the connection with no structural change, shown in Fig.\ \ref{fig:algorithm} (c), are $D_1=1$, $D_e = 0$ $(e \ge 2)$, and the EE and the truncated rDM weight are zero, which should be minimum among the three possible connections.
As a result, the connection of Fig.\ \ref{fig:algorithm} (c) will be selected, and the dimensions of the bonds $a$, $b$, and $e$ remain unity.
(If the stochastic selection is adopted, there is a possibility that the other structure will be selected, but the original structure is most likely to be selected even in this case.)
In such a way, the calculation can be stuck in the TTN containing the isometry whose three legs represent only one basis.
In the present study, we encountered the problem in some cases of the calculations for the Richardson model.
For such cases, we avoided the problem by taking an exceptional treatment in the preparation of the initial TTN.
See Sec.\ \ref{subsec:Rch} and Appendix\ \ref{app:initialTTN} for the details.

\section{Numerical results}\label{sec:results}

\subsection{Types of Calculation and Quantities Computed}\label{subsec:scheme}

We have applied the algorithm discussed in the previous section to the random XY-exchange model under random magnetic fields and the Richardson model.
In the calculation, we can choose (i) the initial TTN, (ii) the scheme to update the local TTN structure, and (iii) the value of $\chi$ with which the structural optimization is performed.
We have carried out various types of calculations specified by (i), (ii), and (iii) and compared them to evaluate their performance.
The types of calculations performed are as follows.

\begin{itemize}
\item[(i)] Initial TTN
\begin{itemize}
\item[$\bullet$] Matrix-Product Network (MPN):
We prepared the MPN as the initial TTN.
The bare spins were arranged one-dimensionally in the order of the site index, and the canonical center is located at the center of the MPN, as shown in Fig.\ \ref{fig:TTNs} (a).
\item[$\bullet$] Perfect-binary tree (PBT):
We prepared the PBT as the initial TTN.
The bare spins were arranged one-dimensionally in the order of the site index, and the canonical center is located at the ``top" of the PBT, as shown in Fig.\ \ref{fig:TTNs} (b).
\item[$\bullet$] TTN constructed by tSDRG method (tSDRG-TTN):
We prepared the initial TTN using the tSDRG, which is a method to construct a TTN suitable to represent the ground state of a system with randomness.
See Ref.\ \cite{SekiHO2020} for the details.
We adopted the gap $\Delta^{\rm I}_{\rm max}$ defined in Ref.\ \cite{SekiHO2020} to determine the order of the isometries to be merged.
This initial TTN constructed by tSDRG was used only for the random XY model.
\end{itemize}
The isometries in the initial TTN were composed of the $\chi$-lowest-energy eigenvectors of the block Hamiltonian of the subsystem consisting of the spins that belong to the incoming bonds of each isometry, as in Ref.\ \cite{HikiharaUOHN2023a}.
For the Richardson model, an exceptional treatment was adopted to avoid the problem of the isometry with three legs having the dimension unity, as discussed in the previous section.
See Appendix\ \ref{app:initialTTN} for details.

\item[(ii)] Scheme to update local TTN structure
\begin{itemize}
\item[$\bullet$] Adopt the local structure with the smallest EE:
We optimized the TTN structure by adopting the local connection that gave the smallest EE among the EEs defined in Eq.\ (\ref{eq:bondEE}).
\item[$\bullet$] Adopt the local structure with the smallest truncated rDM weight\cite{LiRYS2022}:
We optimized the TTN structure by adopting the local connection that gave the smallest truncated rDM weight among the ones defined in Eq.\ (\ref{eq:rDMweight_cut}).
If there were multiple local connections with the weight $\delta\rho$ smaller than $10^{-11}$, we judged that their difference was negligible and adopted the local connection that gave the smallest EE among them.
\item[$\bullet$] Stochastic selection:
We optimized the TTN structure by selecting the local connection stochastically according to the probability $P^{(1)}$ defined by Eq.\ (\ref{eq:stochas1}).
For the effective inverse temperature, we employed the exponential form,
\begin{eqnarray}
\beta = \beta_0 ~ 2^{n/n_\tau},
\label{eq:inv_temp}
\end{eqnarray}
where $n$ was the sweep number and $(\beta_0, n_\tau)$ were control parameters.
\end{itemize}

In addition, we have performed the calculation without structural optimization for comparison.
In this case, the TTN structure was fixed to that of the initial TTN, and only the isometries were improved.
This calculation is equivalent to the conventional variational TTN calculation.

We note that, for the calculation of the stochastic selection, one may employ the same scheme as the above but using the probability $P^{(2)}$ in Eq.\ (\ref{eq:stochas2}) instead of $P^{(1)}$.
We have also performed the calculation with $P^{(2)}$ and found that the results of both the stochastic calculations with $P^{(1)}$ and $P^{(2)}$ coincide semiquantitatively.
Therefore, in the following arguments, we discuss only the results of the calculation with $P^{(1)}$.
The results of the stochastic calculation using $P^{(2)}$ are presented in Appendix\ \ref{app:Stcs2}.

\item[(iii)] $\chi$ for the structural optimization\\
For the calculations with structural optimization, we first performed the structural optimization with $\chi=\chi_{\rm opt}$ to determine the optimal TTN structure.
After that, the isometries were further improved by increasing $\chi$ for the TTN with the structure fixed to the optimized one.
In the actual calculation for the random XY model (the Richardson model), we performed the structural optimization at $\chi_{\rm opt}=16$ or $32$ ($20$ or $40$), and then continued the calculation to optimize isometries by increasing $\chi$ up to $40$ ($80$).
The details, including the number of sweeps and the criteria for judging convergence, are given in Appendix\ \ref{app:Detail_Calcu}.
\end{itemize}

At each step of the calculation, we computed the lowest energy eigenvalue $E_0$ of the effective Hamiltonian $\tilde{\mathcal{H}}$ constructed for the central area.
We then adopted the lowest $E_0$ obtained within a sweep as the variational energy at the sweep.
Here, it should be noted that $E_0$ is the eigenvalue of the effective Hamiltonian expressed in the Hilbert space expanded by the four bonds $a, b, c$, and $d$ in Fig.\ \ref{fig:algorithm}(b), and the corresponding eigenstate $\Psi(abcd)$, that is expressed as Eq.\ (\ref{eq:SVD}) is not strictly the same as the TTN state since the center bond does not suffer from the truncation effect.
Therefore, as a wave function corresponding to the TTN state in which the dimensions of all bonds are upper bounded by $\chi$, we constructed the truncated wave function as
\begin{subequations}
\label{eq:trunc}
\begin{align}
&|\tilde{\Psi}\rangle = \sum_{abcd} \tilde{\Psi}(abcd) |abcd\rangle,
\label{eq:trunc_ket} \\
&\tilde{\Psi}(abcd) = \sum_{e=1}^\chi V_{ab}^e D_e V_{cd}^e,
\label{eq:trunc_wf}
\end{align}
\end{subequations}
for the case that the local connection $(ab|cd)$ [Fig.\ \ref{fig:algorithm} (c)] was selected, and in similar ways for the other connections.
We then calculated the expectation value of the effective Hamiltonian on the truncated state as
\begin{eqnarray}
E'_0 = \frac{\langle \tilde{\Psi} | \tilde{\mathcal{H}} | \tilde{\Psi}\rangle}{\langle \tilde{\Psi} | \tilde{\Psi}\rangle}.
\label{eq:ene_trunc}
\end{eqnarray}
[Note that the truncated wave function $|\tilde{\Psi}\rangle$ defined in Eq.\ (\ref{eq:trunc}) is not strictly normalized.]
We will analyze $E'_0$ in the same way as $E_0$.

\subsection{Random XY-exchange Model Under Random Magnetic Fields}\label{subsec:random}

We have applied the TTN structural optimization algorithm to the spin-1/2 model with random XY-exchange interactions under random magnetic fields.
The model Hamiltonian is given by
\begin{eqnarray}
\mathcal{H}_{\rm r} = \frac{J_{\rm r}}{\sqrt{N}} \sum_{i=1}^{N-1} \sum_{j=i+1}^N
\epsilon_{ij} \left( S^x_i S^x_j + S^y_i S^y_j \right)
- \sum_{i=1}^N h_i S^z_i,
\nonumber \\
\label{eq:Ham_rXY}
\end{eqnarray}
where $J_{\rm r}$ is a parameter to control the strength of the exchange interactions and  ${\bm S}_i = (S^x_i, S^y_i, S^z_i)$ is the spin-1/2 operator at the $i$ th site.
$\epsilon_{ij}$ and $h_i$ are random variables obeying uniform distributions in the range $[0,1]$ and $\left[ -\frac{1}{2}, \frac{1}{2}\right]$, respectively.
We have performed the calculations for the case of the exchange constant $J_{\rm r}=8$ and the system size $N=64$.
In the calculations, we have prepared $\mathcal{N}_{\rm s} = 200$ random samples and computed the lowest-energy state in the subspace with zero total magnetization, $\sum_i S^z_i = 0$.
Note that this state may not be the ground state of each random sample, depending on the random model parameters $\{ \epsilon_{ij} \}$ and $\{ h_i \}$.

We have performed the calculation employing MPN, PBT, and tSDRG-TTN 
as the initial TTN structures.
In the calculation without structural optimization, we have optimized isometries in the TTN with the fixed network structure by sequentially increasing $\chi$ from $16$ to $40$.
For the calculations with structural optimization, two cases of $\chi_{\rm opt}$ were tested: we first performed the TTN structural optimization with $\chi=\chi_{\rm opt}=16$ or $32$.
Then, we fixed the TTN structure and optimized only isometries by increasing $\chi$ up to $40$.
See Appendix\ \ref{app:Detail_Calcu} for details.
Thus, the type of calculation is specified by three conditions:
the initial TTN, optimization scheme, and $\chi_{\rm opt}$ ($=16$ or $32$) with which the TTN structural optimization was performed.
For the stochastic structural optimization, we have performed two runs of the calculation with the different annealing parameters, $(\beta_0, n_\tau)=(0.1, 2), (0.1,4)$, in Eq.\ (\ref{eq:inv_temp}).
Then, among them, we have adopted the result of the run that gives the smaller variational energy $E_0$ ($E'_0$) at $\chi=40$ for the analysis of the random average of $\delta r$ ($\delta r'$) defined below.

To evaluate the performance of each type of calculation, we explore the relative reduction in the variational energy with respect to the energy obtained by the calculation of the fixed structure of MPN at $\chi=16$, 
\begin{eqnarray}
\delta r(\nu, \chi)
&=& \frac{E_0(\nu, \chi)-E_{\rm MPN}(\nu, 16)}{|E_{\rm MPN}(\nu, 16)|},
\label{eq:rltvge_XY} \\
\delta r'(\nu, \chi)
&=& \frac{E'_0(\nu, \chi)-E'_{\rm MPN}(\nu, 16)}{|E'_{\rm MPN}(\nu, 16)|},
\label{eq:rltve1_XY}
\end{eqnarray}
 where $E^{(')}_0(\nu, \chi)$ is the variational energy obtained at the bond dimension $\chi$ for the random sample $\nu = 1, ..., \mathcal{N}_{\rm s}$ and $E^{(')}_{\rm MPN}(\nu, 16)$ is the variational energy of MPN at $\chi=16$.
Then, we examine the random averages of $\delta r(\nu, \chi)$ and $\delta r'(\nu, \chi)$, respectively denoted as $[\delta r(\chi)]$ and $[\delta r'(\chi)]$.
($[ \cdots ]$ denotes the average over $\mathcal{N}_{\rm s}$ random samples.)
We note that $\delta r^{(')}(\nu,\chi)$ can quantitatively evaluate the degree of reduction in the variational energy $E^{(')}(\nu,\chi)$ in each random sample, and therefore, its random average can be a useful indicator of the accuracy improvement of each type of calculation.
In the following, we present only the results of $[\delta r(\chi)]$ since we have found that $[\delta r(\chi)]$ and $[\delta r'(\chi)]$ lead to essentially the same conclusions.

\begin{figure*}
\begin{center}
\includegraphics[width = 49 mm]{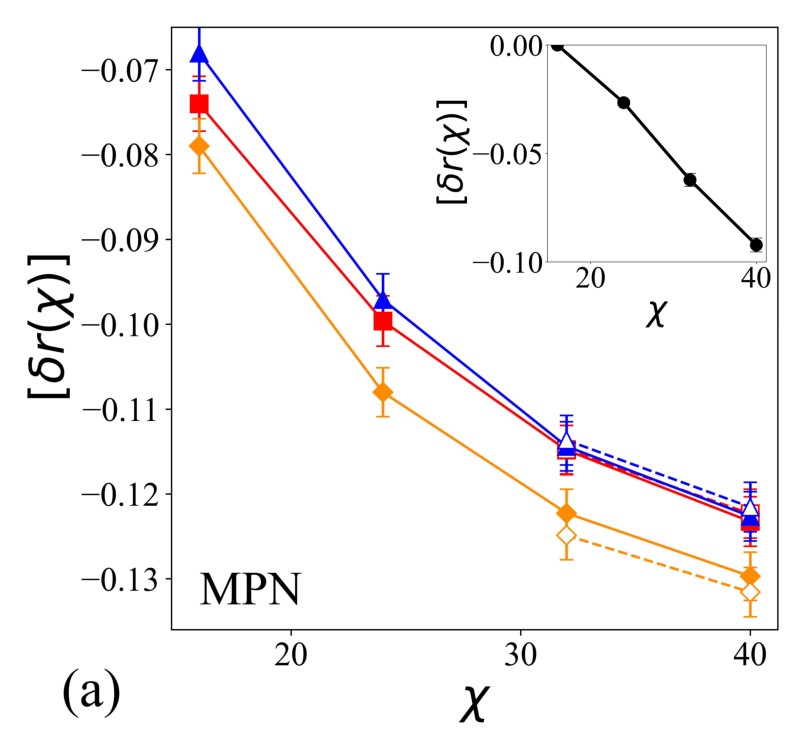}
\includegraphics[width = 48 mm]{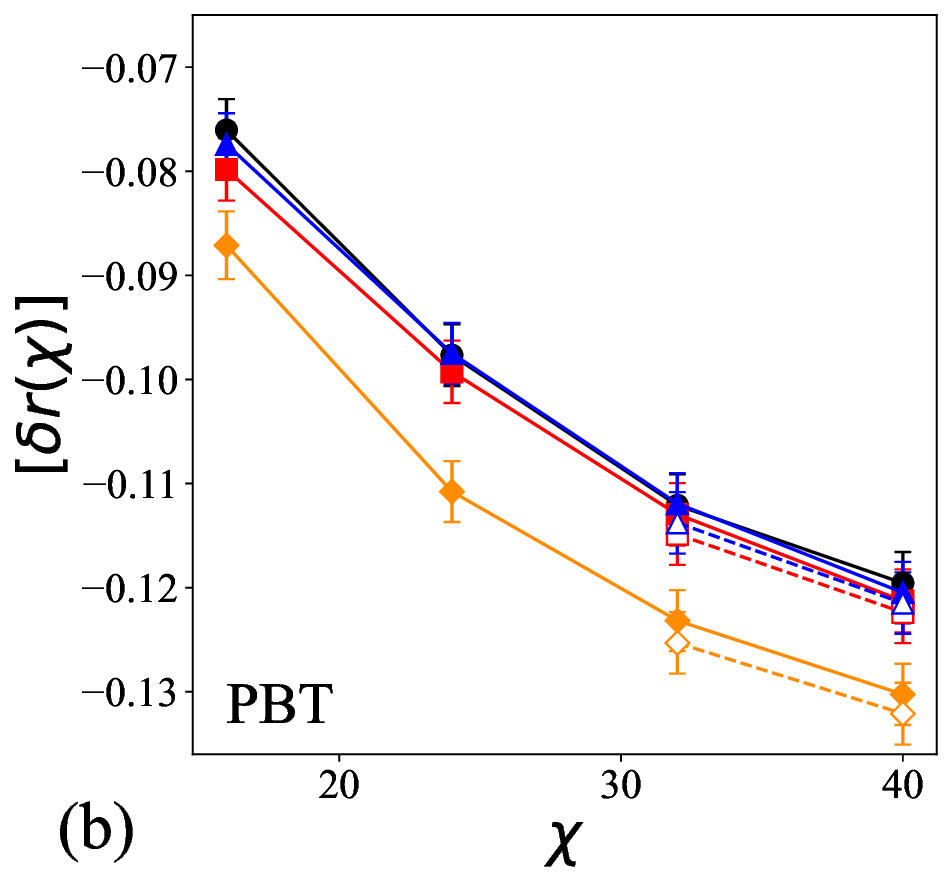}
\includegraphics[width = 74 mm]{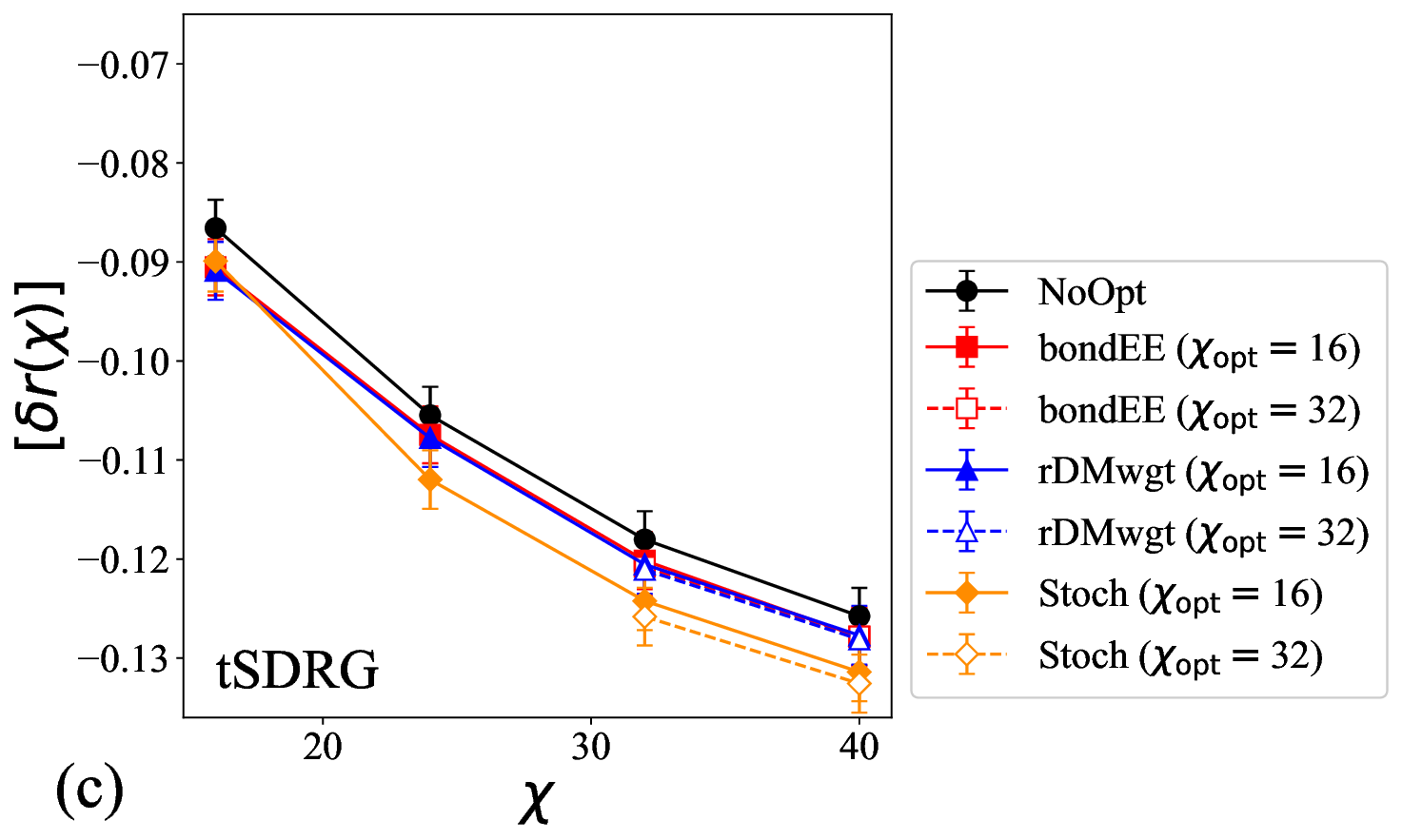}
\caption{
Random averages of the relative reduction in the variational energy, $[\delta r(\chi)]$, for the calculations where the initial TTN is (a) MPN, (b) PBT, and (c) tSDRG-TTN.
``NoOpt" represents the calculation without structural optimization, whereas ``bondEE", ``rDMwgt", and ``Stoch" represent respectively the optimization using the bond EE, that using the truncated rDM weight, and the stochastic optimization.
In (a), the data of the calculation without structural optimization for MPN are shown in the inset.
}
\label{fig:rltvge_XY}
\end{center}
\end{figure*}

\begin{figure}
\begin{center}
\includegraphics[width = 80 mm]{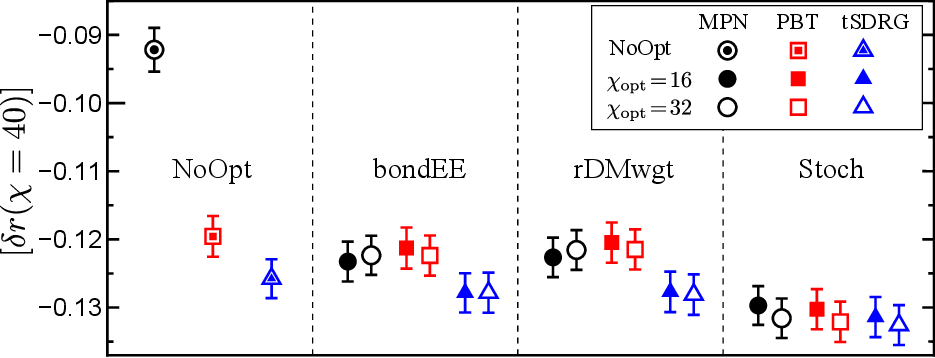}
\caption{
Random averages of the relative reduction in the variational energy, $[\delta r(\chi)]$, at $\chi=40$.
Circles, squares, and triangles represent the results for the calculations where the initial TTN is MPN, PBT, and tSDRG-TTN, respectively.
``NoOpt" represents the calculation without structural optimization, whereas ``bondEE", ``rDMwgt", and ``Stoch" represent respectively the optimization using the bond EE, that using the truncated rDM weight, and the stochastic optimization.
For the calculations with structural optimization, the results for $\chi_{\rm opt}=16$ ($32$) are shown by solid (open) symbols.
}
\label{fig:rltvge_m40_XY}
\end{center}
\end{figure}

Figure\ \ref{fig:rltvge_XY} shows the random average $[\delta r(\chi)]$ as a function of $\chi$.
Also, the values of $[\delta r(\chi=40)]$ for all types are plotted in Fig.\ \ref{fig:rltvge_m40_XY}.
The magnitude of the error bar for the random average $[\delta r(\chi)]$ shown in the figure is obtained as
\begin{eqnarray}
\sigma_r = \sqrt{\frac{1}{\mathcal{N}_{\rm s}(\mathcal{N}_{\rm s}-1)}
\sum_{\nu=1}^{\mathcal{N}_{\rm s}} \left\{ \delta r(\nu, \chi) - [\delta r(\chi)] \right\}^2}.
\nonumber \\
\label{eq:sigma_rltvge_XY}
\end{eqnarray}

In the results without structural optimization, MPN has the poorest accuracy in variational energies, whereas PBT and tSDRG-TTN yield a relatively good accuracy.
As for the results of non-stochastic structural optimization with the bond EE [Eq.\ (\ref{eq:bondEE})] and the truncated rDM weight [Eq.\ (\ref{eq:rDMweight_cut})], a significant improvement is observed when the initial TTN is MPN, reflecting the poor accuracy of the MPN calculation without structural optimization.
When the initial TTN is PBT or tSDRG-TTN, there is an improvement in accuracy although it is small.
Besides, when the initial TTN is MPN and PBT, the accuracy in the variational energies is worse than that of tSDRG-TTN without structural optimization.
These results suggest that the results of non-stochastic structural optimization are under the influence of the initial TTN.
(Incidentally, this observation indicates the effectiveness of tSDRG for the random XY model.)
We note that there is no notable difference between the results of the structural optimizations with the bond EE and the truncated rDM weight; their performance is comparable, at least for the random XY model and $\chi_{\rm opt}$ treated.
For the stochastic structural optimization, the variational energy is further reduced, yielding the best accuracy among the results obtained here.
Moreover, it is important to note that the results of the stochastic optimization achieve almost the same level of accuracy, independent of the initial TTN.
These results suggest that the stochastic structural optimization works well for the random XY model and succeeds in escaping the influence of the initial TTN.

Concerning the effect of $\chi_{\rm opt}$ at which the structural optimization is performed, the results of $\chi_{\rm opt}=32$ yield slightly lower variational energy than those of $\chi_{\rm opt}=16$ for the stochastic optimization.
However, the difference is within the statistical error and is basically smaller than the difference from the results of non-stochastic optimizations.
This observation indicates that the value of $\chi_{\rm opt}$ has only a minor effect compared to the initial TTN and the updating scheme, at least within the range of $\chi$ we treated here.

Next, we discuss the EEs in the TTN.
Since the EE on the physical bonds, the bonds directly connected to the bare spins, takes the same value regardless of the TTN structure, 
we consider only the EEs on the auxiliary bonds in the following analysis.
In the analysis, we use the data of EEs obtained in the sweep which yields the lowest variational energy $E_0$ at each $\chi$.
For the stochastic structural optimization, we use the EEs obtained by the run that yields the lower variational energy $E_0$ at $\chi=40$ between the two runs with different annealing parameters.

\begin{figure}
\begin{center}
\includegraphics[width = 80 mm]{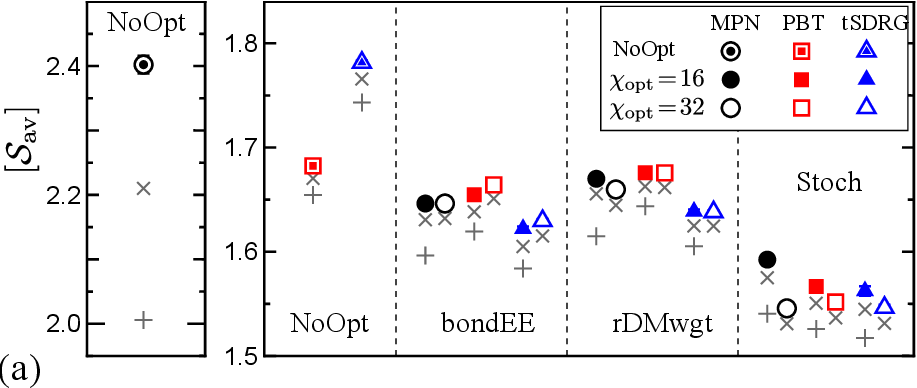}
\includegraphics[width = 80 mm]{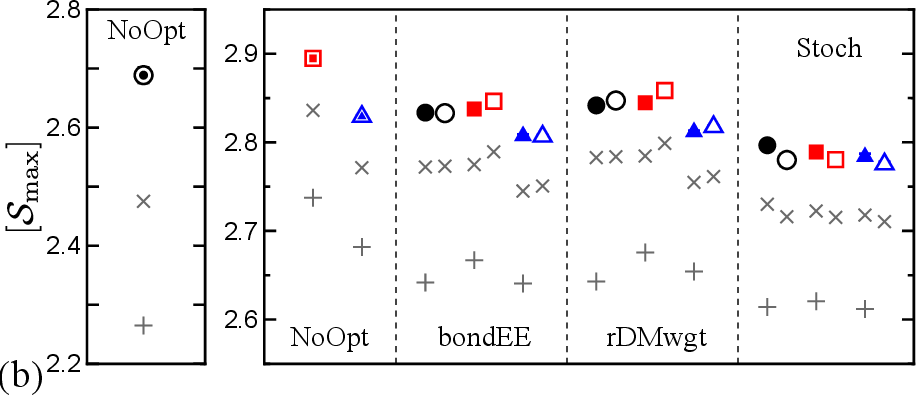}
\caption{
(a) $\left[ \mathcal{S}_{\rm av}\right]$ and (b) $\left[ \mathcal{S}_{\rm max}\right]$, that are respectively the random averages of the average values $\mathcal{S}_{\rm av}(\nu)$ and maximum values $\mathcal{S}_{\rm max}(\nu)$ of EEs on auxiliary bonds in each random sample.
Circles, squares, and triangles represent the results at $\chi=40$ for the calculations where the initial TTN is MPN, PBT, and tSDRG-TTN, respectively.
``NoOpt" represents the calculation without structural optimization, whereas ``bondEE", ``rDMwgt", and ``Stoch" represent respectively the optimization using the bond EE, that using the truncated rDM weight, and the stochastic optimization.
For the calculations with structural optimization, the results for $\chi_{\rm opt}=16$ ($32$) are shown by solid (open) symbols.
Grey crosses represent the results at $\chi=32$ ($\times$) and $\chi=24$ ($+$); the latter is not present for the structural optimizations with $\chi_{\rm opt}=32$.
The results for the calculation without structural optimization for MPN are plotted in a different scale.
Error bars are plotted only for the results at $\chi=40$, although they are smaller than the symbols in most cases.
}
\label{fig:EEavmax_XY}
\end{center}
\end{figure}

To see how the EEs in the TTN are suppressed by the structural optimization, we explore the average and maximum values of EE.
Figure\ \ref{fig:EEavmax_XY} shows the random averages of $\mathcal{S}_{\rm av}(\nu)$ and $\mathcal{S}_{\rm max}(\nu)$, $\left[ \mathcal{S}_{\rm av}\right]$ and $\left[ \mathcal{S}_{\rm max}\right]$, where $\mathcal{S}_{\rm av}(\nu)$ and $\mathcal{S}_{\rm max}(\nu)$ are respectively the average and maximum values of EE on the auxiliary bonds in each random sample.
Error bars are obtained by a formula similar to Eq.\ (\ref{eq:sigma_rltvge_XY}).
We note that since $\left[ \mathcal{S}_{\rm av}\right]$ and $\left[ \mathcal{S}_{\rm max}\right]$ are increasing functions of $\chi$, one should analyze them at a sufficiently large $\chi$ to compare the TTNs obtained by different types of structural optimization.

For the average EE, $\left[ \mathcal{S}_{\rm av}\right]$, the result for MPN without structural optimization is remarkably large and has a pronounced $\chi$ dependence, while the $\chi$ convergence is almost achieved at $\chi=40$ for the other data.
This observation indicates that in MPN, there is a significant fraction of auxiliary bonds whose EEs are too large to be borne even by a bond with $\chi=40$.
The non-stochastic structural optimizations with the bond EE and the truncated rDM weight reduce $\left[ \mathcal{S}_{\rm av}\right]$ to some extent, but the reductions are modest and their dependence on the initial TTN is visible, indicating that the results are still under the influence of the initial TTNs.
In contrast, $\left[ \mathcal{S}_{\rm av}\right]$ obtained by the stochastic structural optimization basically achieves small values independently of the initial TTNs.
This result demonstrates the effectiveness of the stochastic optimization for the random XY model.
As for the dependence of the results on $\chi_{\rm opt}$, the difference between $\left[ \mathcal{S}_{\rm av}\right]$ for $\chi_{\rm opt}=16$ and $\chi_{\rm opt}=32$ is basically small compared to the difference due to the initial TTN or the updating scheme, indicating that the effect of $\chi_{\rm opt}$ is secondary.

The data of $\left[ \mathcal{S}_{\rm max}\right]$ suffer from relatively large $\chi$ dependence; however, they lead us to the following conclusions that are basically consistent with the ones we read from the data of $[\mathcal{S}_{\rm av}]$. 
That is, the results of the calculation without structural optimization exhibit specific behaviors reflecting the characteristics of each TTN.
[We note that $[\mathcal{S}_{\rm max}]$ for the fixed structure of MPN at $\chi=40$ is smaller than those for other calculations, but its large $\chi$ dependence implies that it may become greater than the others at $\chi \to \infty$.]
The results of the non-stochastic schemes exhibit a sizable dependence on the initial TTN.
The results of stochastic structural optimization yield the smallest $\left[ \mathcal{S}_{\rm max}\right]$ [except the one for the fixed MPN], with small dependence on the initial TTN.
In addition, compared to the differences by the initial TTN and structural optimization scheme, the differences by $\chi_{\rm opt}$ are minor.

\begin{figure*}
\begin{center}
\includegraphics[width = 140 mm]{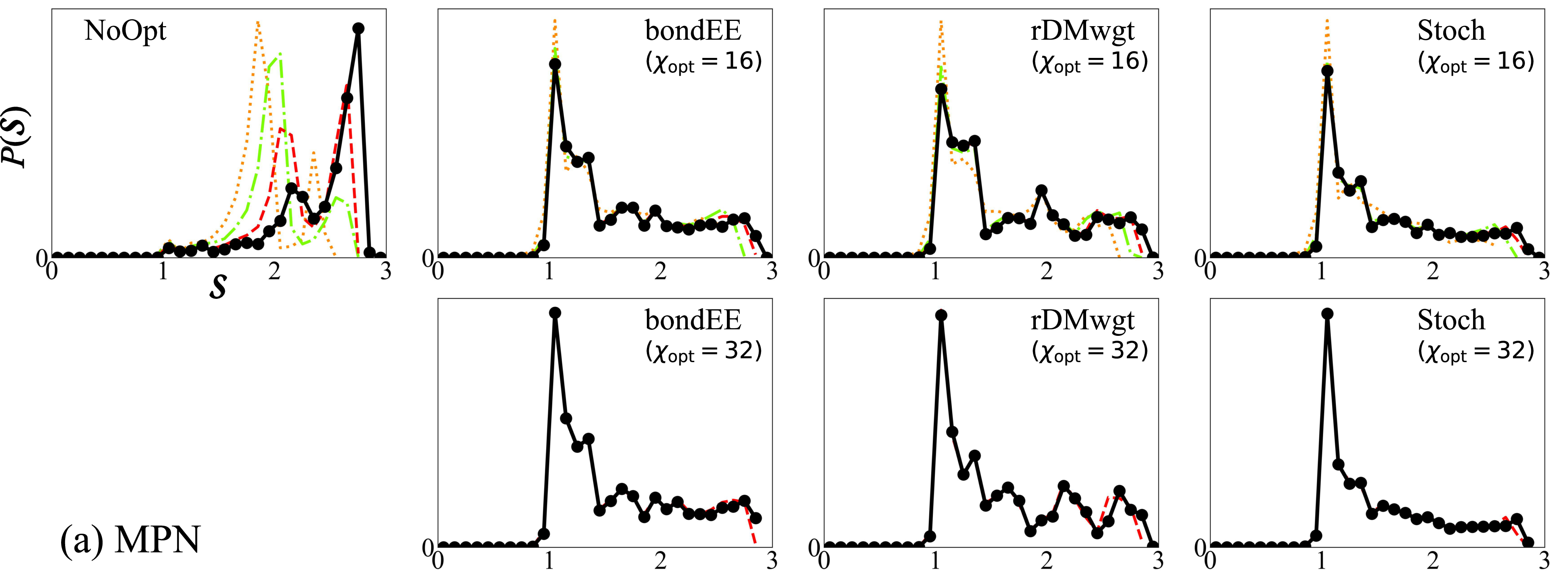}
\includegraphics[width = 140 mm]{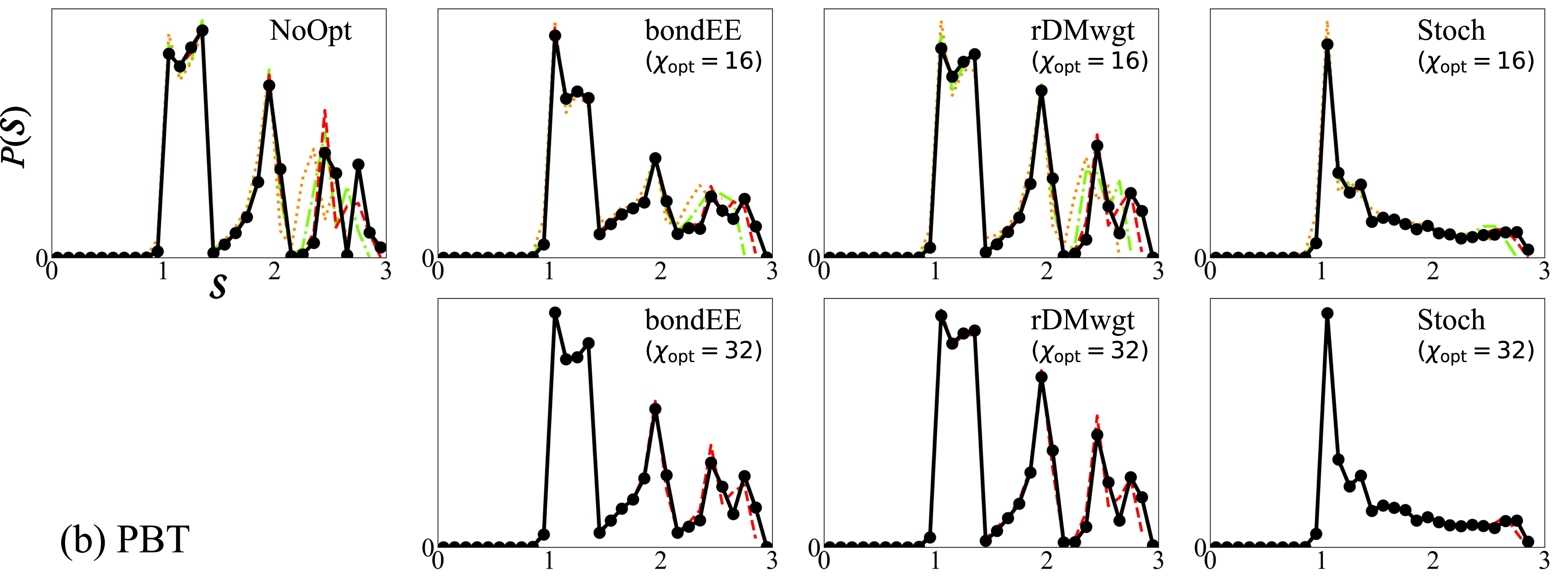}
\includegraphics[width = 140 mm]{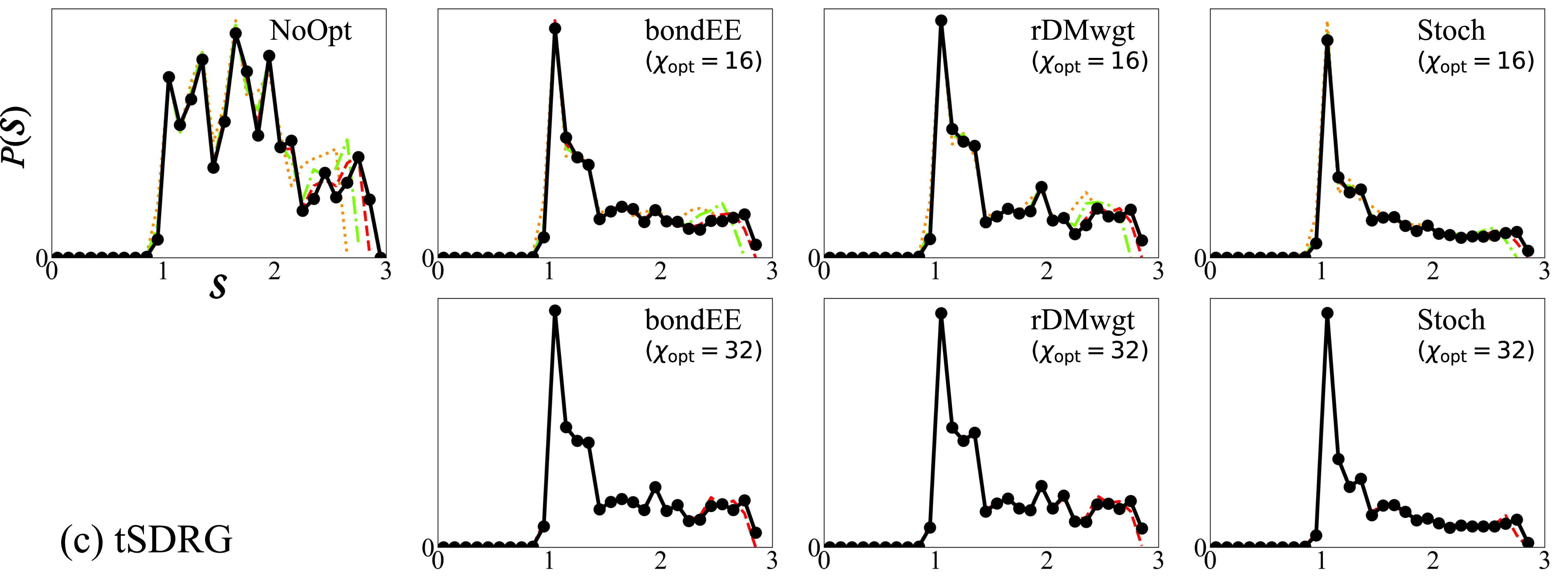}
\caption{
Histogram $P(\mathcal{S})$ of entanglement entropy $\mathcal{S}$ on auxiliary bonds in TTNs obtained by the calculations where the initial TTN is (a) MPN, (b) PBT, and (c) tSDRG-TTN, for $\mathcal{N}_{\rm s} = 200$ random samples.
``NoOpt" represents the calculation without structural optimization, whereas ``bondEE", ``rDMwgt", and ``Stoch" represent respectively the optimization using the bond EE, that using the truncated rDM weight, and the stochastic optimization.
Circles connected by solid lines represent the data at $\chi=40$.
Dashed (red), dot-dashed (green), and dotted (orange) lines are the data at $\chi=32, 24$, and $16$, respectively.
The size of bins is set to $\Delta \mathcal{S} = 0.1$.
}
\label{fig:EEhist_XY}
\end{center}
\end{figure*}

We also analyze the histogram of EEs obtained by each type of calculation.
We calculate the histogram $P(\mathcal{S})$ using the EEs on the auxiliary bonds in all random samples [the total number of EEs is $\mathcal{N}_{\rm s} (N-3)$], where the bin size is set to $\Delta \mathcal{S} = 0.1$.
Figure\ \ref{fig:EEhist_XY} presents the obtained histograms.
The $\chi$ dependence is sizable only in the histogram of the calculation without structural optimization for MPN and irrelevant for the other results.
The histograms for the scheme without structural optimization have a different shape depending on the initial TTNs, which should reflect the characteristics of each TTN.
When applied to PBT, the non-stochastic optimizations yield histograms with a similar shape to that of the calculation without structural optimization, suggesting that the TTN structures obtained there remain under the influence of the initial PBT.
Also, when the initial TTN is MPN or tSDRG-TTN, the histograms obtained by the non-stochastic optimizations still exhibit peculiar characteristics such as bumpy behavior in the region of large EE ($1.2 \lesssim \mathcal{S} \lesssim 3.0$), although their overall structures tend to approach the histograms obtained by the stochastic optimization.
In contrast, the histograms obtained by the stochastic optimization converge to nearly the same shape regardless of the initial TTN;
$P(\mathcal{S})$ exhibits a sharp peak around $\mathcal{S} \sim 1.0 - 1.1$, and from the peak, it decreases with a gentle tail as $\mathcal{S}$ increases.
These histograms should be the ones of the best TTNs achieved in the present results.
Finally, no essential difference is found between the results for $\chi_{\rm opt}=16$ and $32$.

Those observations on the EEs are consistent with the findings obtained from the analyses of the variational energies.
We thereby conclude that the stochastic structural optimization works well for the random XY model and succeeds in obtaining the best TTN structure yielding the lowest variational energy independently of the initial TTN.

\subsection{Richardson Model}\label{subsec:Rch}

We have applied the TTN structural optimization algorithm to the Richardson model, that is a spin-1/2 model containing all-to-all XY-exchange interactions under magnetic fields.
The Hamiltonian is given by
\begin{eqnarray}
\mathcal{H}_{\rm R} = \frac{J_{\rm R}}{N}
\sum_{i=1}^{N-1} \sum_{j=i+1}^N (S^x_i S^x_j + S^y_i S^y_j )
- \sum_{i=1}^N h_i S^z_i,
\label{eq:Ham_Rch}
\end{eqnarray}
where $J_{\rm R}$ is the exchange-coupling parameter.
The site-dependent longitudinal magnetic field is taken as
\begin{eqnarray}
h_i = \frac{i-1}{N-1} - \frac{1}{2},
\label{eq:hzi_Rch}
\end{eqnarray}
so that the field $h_i$ linearly increases from $h_1=-\frac{1}{2}$ to $h_N = \frac{1}{2}$ with the site index $i$.
The Richardson model [Eq.\ (\ref{eq:Ham_Rch})] is known to be exactly solvable and has been studied for many years in condensed-matter physics, particularly in connection with the BCS superconductivity, and nuclear physics.\cite{DelftR2001,AsoreyFS2002,DukelskyPS2004}

As in the case of the random XY model in the previous section, we have performed several types of the structural optimization calculations to obtain the lowest-energy state in the subspace of zero total magnetization $\sum_i S^z_i = 0$.
The exchange constant and system size were $J_{\rm R} = 1, 4, 16, 64$ for $N=64$ and $J_{\rm R}=1$ for $N=128$.
The maximum bond dimension was $\chi=80$.
Here, we note that the system size $N$ and the bond dimension $\chi$ are much larger than those for the random XY model.
This is because the computational cost of the exchange interaction terms for the Richardson model is much smaller (roughly speaking, $1/N$ times smaller) than that for the random XY model, reflecting the property of the Richardson model that all spin pairs are coupled by the exchange interaction of the same strength.

We have performed the calculations starting from the initial TTN of MPN and PBT.
In the calculations without structural optimization, the bond dimension was increased from $\chi=20$ to $80$.
In the calculations with structural optimization, the TTN structure was optimized at $\chi=\chi_{\rm opt}=20$ or $40$, and then, the calculation to optimize only the isometries was done with $\chi$ up to 80.
See Appendix\ \ref{app:Detail_Calcu} for the details.
For the stochastic optimization, we have performed the calculations of four runs, i.e., two runs (with different random seeds) for each set of the annealing parameters $(\beta_0, n_\tau) = (0.1, 2)$ and $(0.1, 4)$.
In addition, in order to avoid the deadlock of the calculation due to an isometry whose all three legs have dimension $\chi=1$ discussed in Sec.\ \ref{sec:algo}, we adopted an exceptional treatment in the preparation process of the initial TTN.
The details of the treatment are presented in Appendix\ \ref{app:initialTTN}.
In our calculations, the treatment was necessitated only for the cases where $(J_{\rm R}, N) = (64, 64)$ and the initial TTN was PBT.

\begin{figure}
\begin{center}
\includegraphics[width = 75 mm]{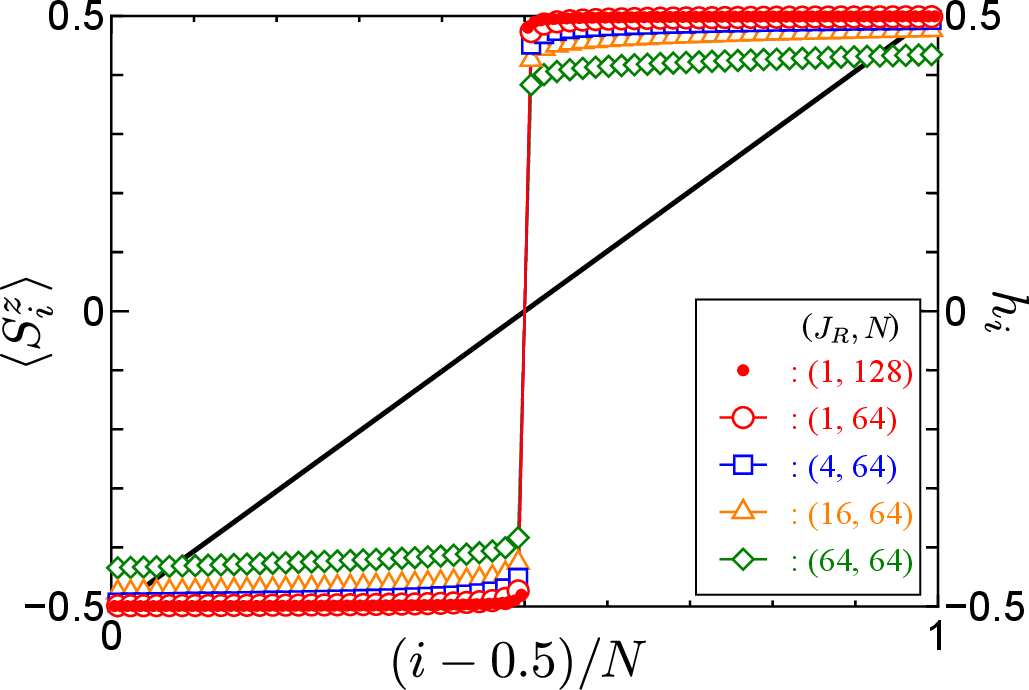}
\caption{
Local spin polarization $\langle S^z_i \rangle$.
The plots show the results obtained by the calculation that yields the smallest relative error $\delta e$ for each model parameter $(J_{\rm R},N)$.
The thick black line represents the magnetic field $h_i$.
}
\label{fig:Szl_Rch}
\end{center}
\end{figure}

Before proceeding to the performance evaluation, let us examine the nature of the lowest energy state to be analyzed.
In the Richardson model, the XY exchange interactions, which entangle all spins to form a large spin cluster, compete with the magnetic fields, which polarize the spins in the $S^z$ direction.
Figure\ \ref{fig:Szl_Rch} presents the data of the spin polarization $\langle S^z_i \rangle$ in the lowest-energy states.
For $J_{\rm R}=1$, where the exchange interaction is relatively weak, most of the spins except for those in the vicinity of $i=N/2$ are strongly polarized, $\langle S^z_i \rangle \sim \pm 1/2$, and almost classically ordered.
Only the spins near $i=N/2$, where the field is weak, have the polarization slightly reduced by the exchange interactions;
For example, $\langle S^z_{65} \rangle = - \langle S^z_{64}\rangle = 0.4801$ for $J_{\rm R}=1$ and $N=128$.
In this situation, the accuracy of the TTN calculation depends on how precisely the TTN state can represent the entanglement among those spins near $i=N/2$.
As the exchange interaction $J_{\rm R}$ increases, the magnitude of the spin polarization decreases gradually, signaling that the spins start to entangle with each other to form a large spin cluster.
A point to be noted here is that the $i$ dependence of the spin polarization remains flat except for the narrow region around $i=N/2$.
Thus, it is expected that the location of such spins with nearly the same polarization in a TTN is irrelevant to the accuracy of the calculation.
(Note that the exchange interactions take the same strength for all spin pairs and have no position dependence.)
This point will be essential in the later discussion on the TTN structure.

In order to discuss the accuracy of the TTN optimization quantitatively, we evaluate the relative error in the variational energies defined by
\begin{eqnarray}
\delta e(\chi) &=& \frac{E_0(\chi)-E_{\rm exact}}{|E_{\rm exact}|},
\label{eq:rltverr_Rch} \\
\delta e'(\chi) &=& \frac{E'_0(\chi)-E_{\rm exact}}{|E_{\rm exact}|},
\label{eq:rltverrp_Rch}
\end{eqnarray}
where $E_{\rm exact}$ is the exact lowest energy obtained from the exact solution for the Richardson model\cite{DelftR2001,AsoreyFS2002,DukelskyPS2004}; See Appendix\ \ref{app:Exact_Rch} for the calculation of $E_{\rm exact}$.
We will see that $\delta e$ and $\delta e'$ exhibit qualitatively similar behavior, although $\delta e'$ is by definition not less than $\delta e$.
However, the difference between $\delta e$ and $\delta e'$ is noticeable for PBT without structural optimization.
This is because in PBT, the EE on the ``top" bond is significantly larger than EEs on other bonds, and $\delta e$ calculated at the top bond can be further reduced by avoiding the effect of the truncation of that large EE.

\begin{figure*}
\begin{center}
\includegraphics[width = 48 mm]{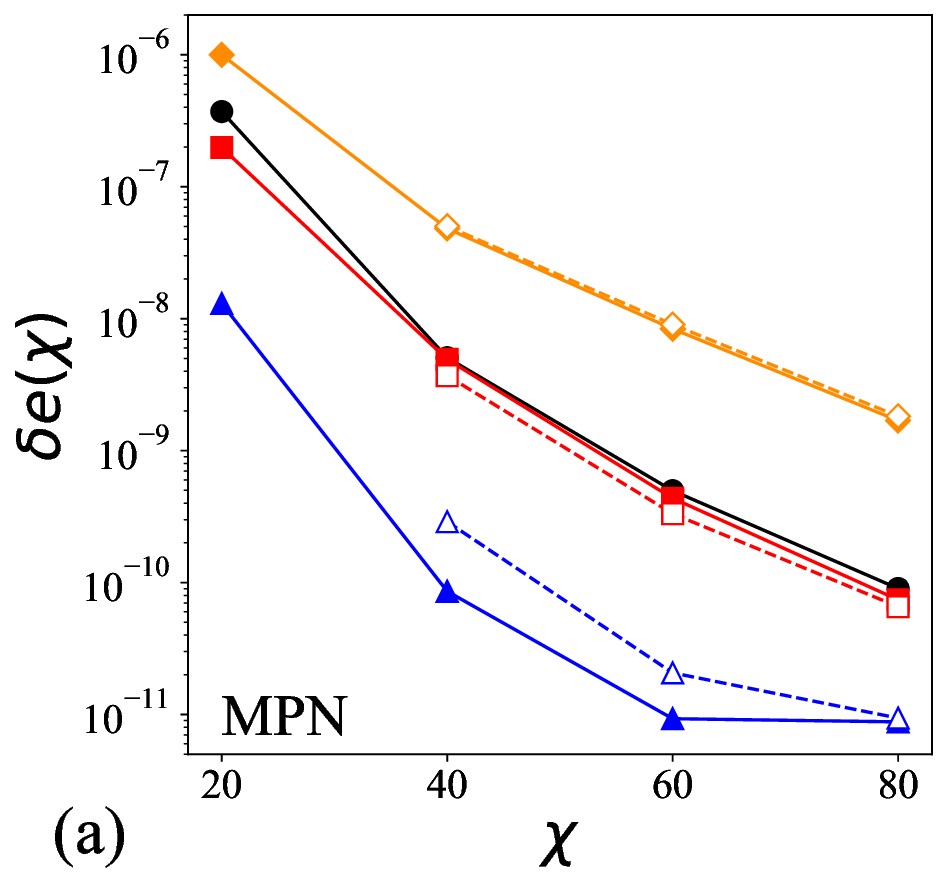}
\includegraphics[width = 74 mm]{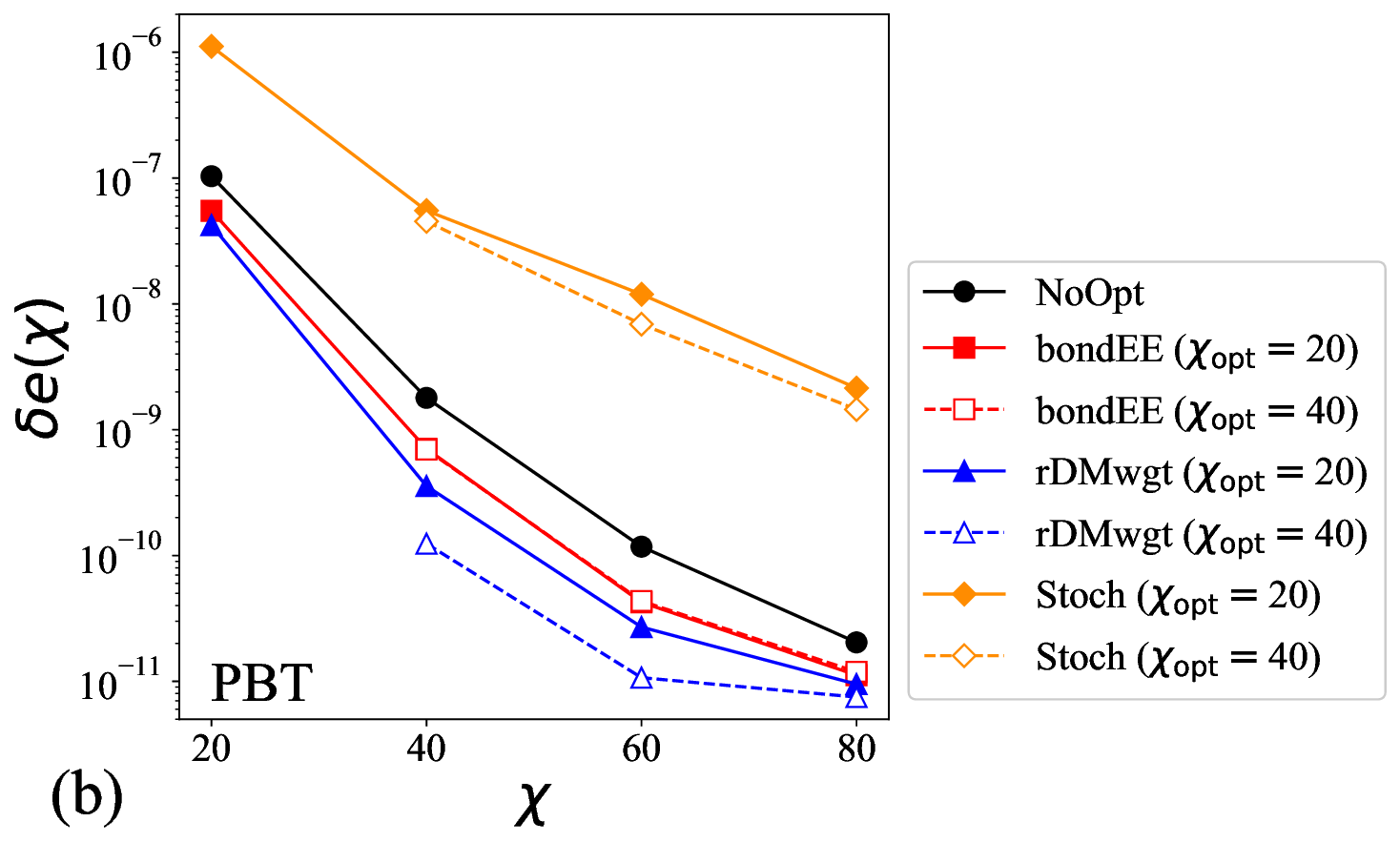}
\caption{
Relative error in the variational energy, $\delta e(\chi)$, for the calculations where the initial TTN is (a) MPN and (b) PBT.
The model parameters are $J_{\rm R}=1$ and $N=128$.
``NoOpt" represents the calculation without structural optimization, whereas ``bondEE", ``rDMwgt", and ``Stoch" represent respectively the optimization using the bond EE, that using the truncated rDM weight, and the stochastic optimization.
For the stochastic optimization, the result of the run yielding the lowest variational energy at $\chi=80$ among the four runs is plotted.
}
\label{fig:rltverr_Rch}
\end{center}
\end{figure*}

\begin{figure}
\begin{center}
\includegraphics[width = 80 mm]{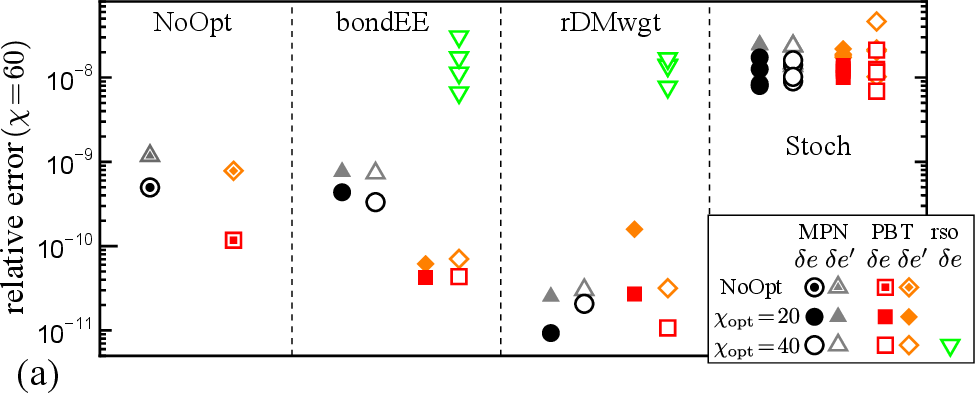}
\includegraphics[width = 77 mm]{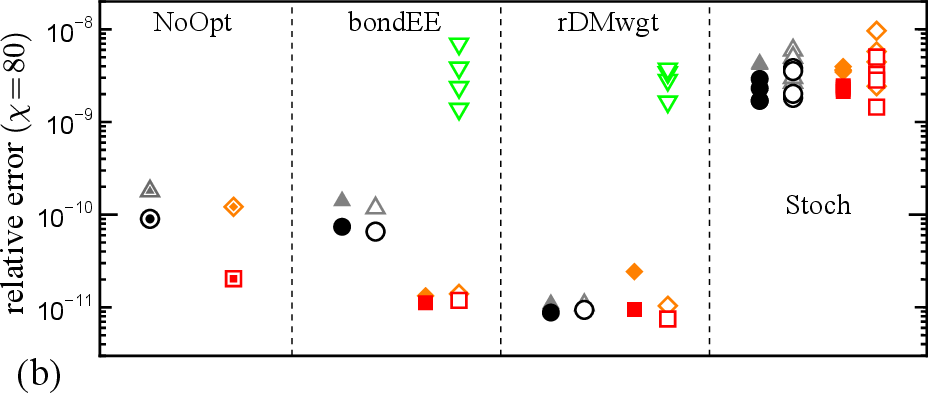}
\caption{
Relative errors in the variational energies, $\delta e(\chi)$ and $\delta e'(\chi)$, at (a) $\chi=60$ and (b) $\chi=80$.
The model parameters are $J_{\rm R}=1$ and $N=128$.
Circles and squares (triangles and diamonds) represent $\delta e(\chi)$ [$\delta e'(\chi)$] for the calculations where the initial TTN is MPN and PBT, respectively.
``NoOpt" represents the calculation without structural optimization, whereas ``bondEE", ``rDMwgt", and ``Stoch" represent respectively the optimization using the bond EE, that using the truncated rDM weight, and the stochastic optimization.
For the calculations with structural optimization, the results for $\chi_{\rm opt}=20$ ($40$) are shown by solid (open) symbols.
For the stochastic optimization, the results of four runs are plotted.
Downward triangles represent $\delta e(\chi)$ obtained by the calculations of non-stochastic structural optimizations with $\chi_{\rm opt}=40$ starting from PBTs with four different random site orderings (rso); See the main text.
For the stochastic optimization and the non-stochastic optimizations with the initial PBTs with random-site orderings, some of the results overlap.
}
\label{fig:rltverr_m6080_Rch}
\end{center}
\end{figure}

Let us compare $\delta e$ ($\delta e'$) of various types of calculation.
Figure\ \ref{fig:rltverr_Rch} shows the $\chi$ dependence of $\delta e(\chi)$ for $(J_{\rm R},N)=(1, 128)$.
We also show in Fig.\ \ref{fig:rltverr_m6080_Rch} the data of $\delta e(\chi)$ and $\delta e'(\chi)$ for $\chi=60$ and $80$.
We find that for the calculations without structural optimization, PBT gives a rather accurate result.
As for the non-stochastic optimization schemes, the optimization using the truncated rDM weight realizes a sizable reduction in the relative error;
$\delta e$ obtained by the optimization for the initial MPN with $\chi_{\rm opt}=20$ converge up to $\delta e(\chi=60)=9.3 \times 10^{-12}$ at $\chi=60$ and $\delta e(\chi=80)=8.8 \times 10^{-12}$ at $\chi=80$, indicating that the calculation even at $\chi=60$ gives a nearly exact result within the accuracy limit of our calculation.\cite{accuracy_limit}
On the other hand, the optimization using the bond EE reduces the relative error compared to that without structural optimization when applied to the initial PBT, while the reduction is almost negligible for the initial MPN.
These observations suggest that the non-stochastic structural optimization schemes are able to realize the accuracy improvement for the Richardson model, although the improvement may occasionally be small.
In contrast, the accuracy of the stochastic structural optimization is poor, even worse than that without structural optimization, indicating that the stochastic optimization does not work well for the Richardson model.

\begin{figure}
\begin{center}
\includegraphics[width = 80 mm]{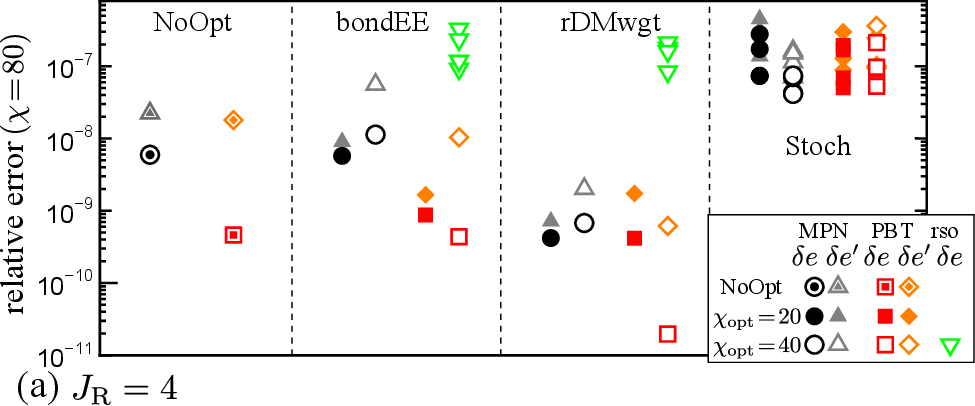}
\includegraphics[width = 77 mm]{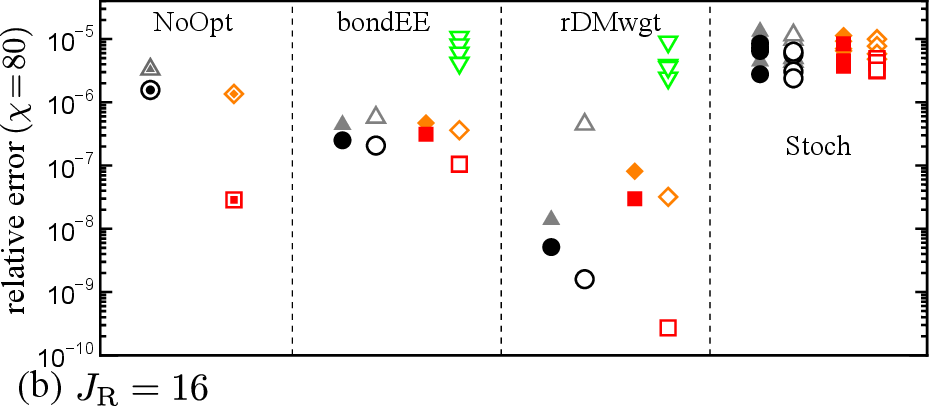}
\includegraphics[width = 77 mm]{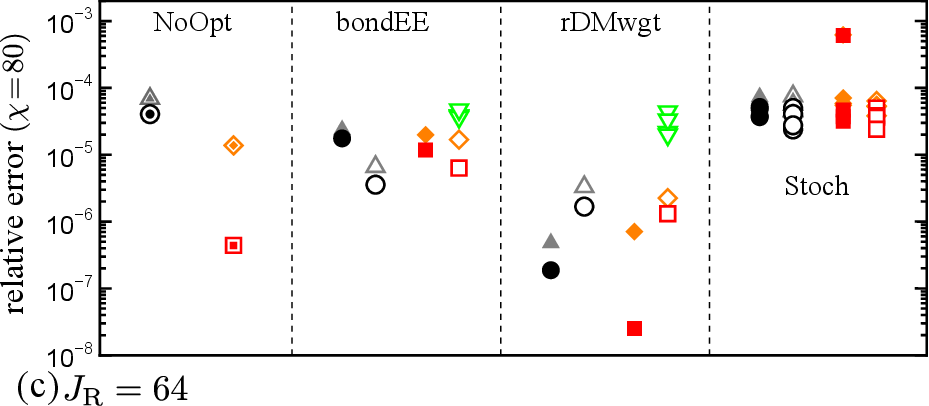}
\caption{
Relative errors in the variational energies, $\delta e(\chi)$ and $\delta e'(\chi)$, at $\chi=80$ for $N=64$ and (a) $J_{\rm R}=4$, (b) $J_{\rm R}=16$, and (c) $J_{\rm R}=64$.
Circles and squares (triangles and diamonds) represent $\delta e(\chi)$ [$\delta e'(\chi)$] for the calculations where the initial TTN is MPN and PBT, respectively.
``NoOpt" represents the calculation without structural optimization, whereas ``bondEE", ``rDMwgt", and ``Stoch" represent respectively the optimization using the bond EE, that using the truncated rDM weight, and the stochastic optimization.
For the calculations with structural optimization, the results for $\chi_{\rm opt}=20$ ($40$) are shown by solid (open) symbols.
For the stochastic optimization, the results of four runs are plotted.
Downward triangles represent $\delta e(\chi)$ obtained by the calculations of non-stochastic structural optimizations with $\chi_{\rm opt}=40$ starting from PBTs with four different random site orderings (rso); See the main text.
For the stochastic optimization and the non-stochastic optimizations with the initial PBTs with random-site orderings, some of the results overlap.
}
\label{fig:rltverr_m80_Rch_J040-640}
\end{center}
\end{figure}

In Fig.\ \ref{fig:rltverr_m80_Rch_J040-640}, we show the results of $\delta e$ and $\delta e'$ for $J_{\rm R}= 4, 16, 64$ and $N=64$.
These data for large $J_{\rm R}$ also exhibit basically the same results as those for $J_{\rm R}=1$.
That is, when the initial TTN is MPN, the non-stochastic optimizations can achieve accuracy improvement, although they are occasionally stuck in a state of poor accuracy.
PBT yields a result with good accuracy as it is, but even better accuracy may be achieved by the optimization using the truncated rDM weight.
In contrast, the stochastic optimization is not accurate and does not work well for the Richardson model.

\begin{figure}
\begin{center}
\includegraphics[width = 80 mm]{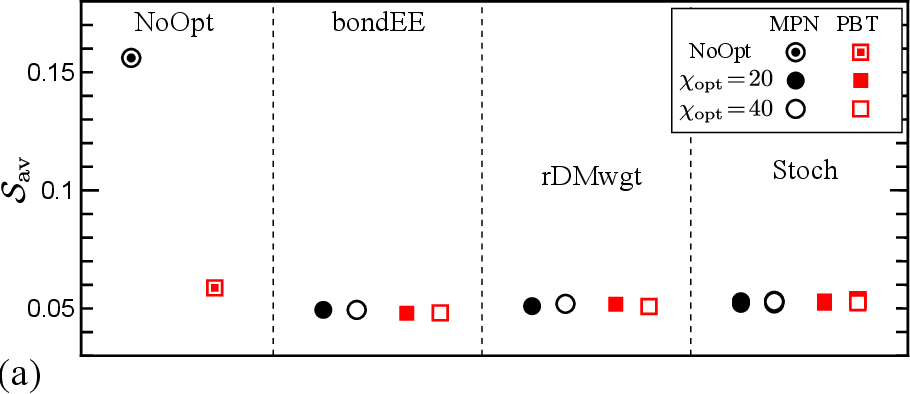}
\includegraphics[width = 80 mm]{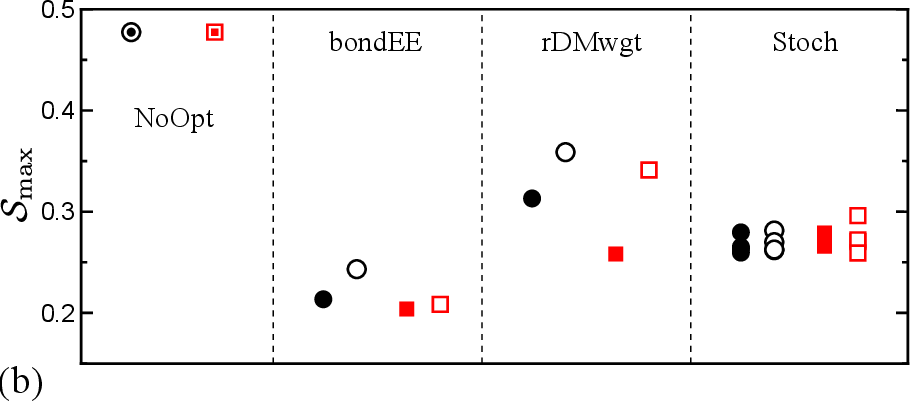}
\caption{
(a) Average value $\mathcal{S}_{\rm av}$ and (b) maximum value $\mathcal{S}_{\rm max}$ of EEs on auxiliary bonds for $J_{\rm R}=1$ and $N=128$.
Circles and squares represent the results at $\chi=80$ of the calculations where the initial TTN is MPN and PBT, respectively.
``NoOpt" represents the calculation without structural optimization, whereas ``bondEE", ``rDMwgt", and ``Stoch" represent respectively the optimization using the bond EE, that using the truncated rDM weight, and the stochastic optimization.
For the calculations with structural optimization, the results for $\chi_{\rm opt}=20$ ($40$) are shown by solid (open) symbols.
For the stochastic optimization, the results of four runs are plotted.
The $\chi$ convergence has been confirmed by comparing the results at $\chi=80$ and $60$.
For the stochastic optimization, some of the results overlap.
}
\label{fig:EEavmax_Rch}
\end{center}
\end{figure}

Figure\ \ref{fig:EEavmax_Rch} shows the average and maximum values of the EEs on auxiliary bonds, $\mathcal{S}_{\rm av}$ and $\mathcal{S}_{\rm max}$, for $J_{\rm R}=1$ and $N=128$.
The average value $\mathcal{S}_{\rm av}$ is remarkably large for the calculation without structural optimization for the initial MPN, while $\mathcal{S}_{\rm av}$ is small for the other cases.
The maximum values $\mathcal{S}_{\rm max}$ for the calculations without structural optimization, where the TTN structures are MPN or PBT, are equal and large, which are the EE between the subsystems of $1 \le i \le N/2$ and $N/2+1 \le i \le N$.
Compared with them, $\mathcal{S}_{\rm max}$ for the other cases are small;
In particular, the optimization using the bond EE yields the smallest $\mathcal{S}_{\rm max}$, indicating that the scheme, which directly uses the EE as a measure in selecting the local TTN structure, is the most suitable for minimizing $\mathcal{S}_{\rm max}$.
The qualitatively same features of $\mathcal{S}_{\rm av}$ and $\mathcal{S}_{\rm max}$ are also observed for other model parameters $J_{\rm R}$ and $N$ (data not shown), while the values of $\mathcal{S}_{\rm av}$ and $\mathcal{S}_{\rm max}$ increase as $J_{\rm R}$ increases.
Here, we recall that the relative errors in the variational energy were smallest for the optimization using the truncated rDM weight in most cases.
These results thus suggest that there was no strong correlation between the accuracy of the variational energy and the average and maximum EEs in the Richardson model.
The observations indicate the importance of choosing an adequate scheme of network optimization depending on the model treated and the quantity to be minimized.

\begin{figure*}
\begin{center}
\includegraphics[width = 140 mm]{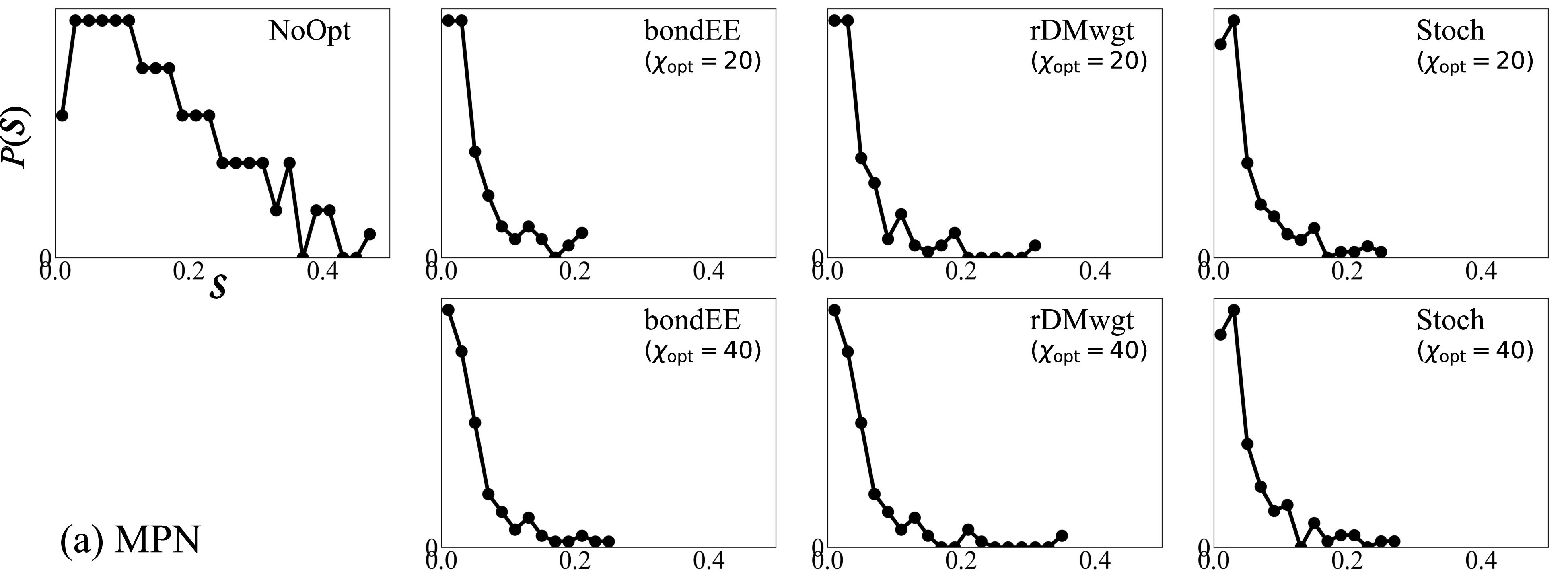}
\includegraphics[width = 140 mm]{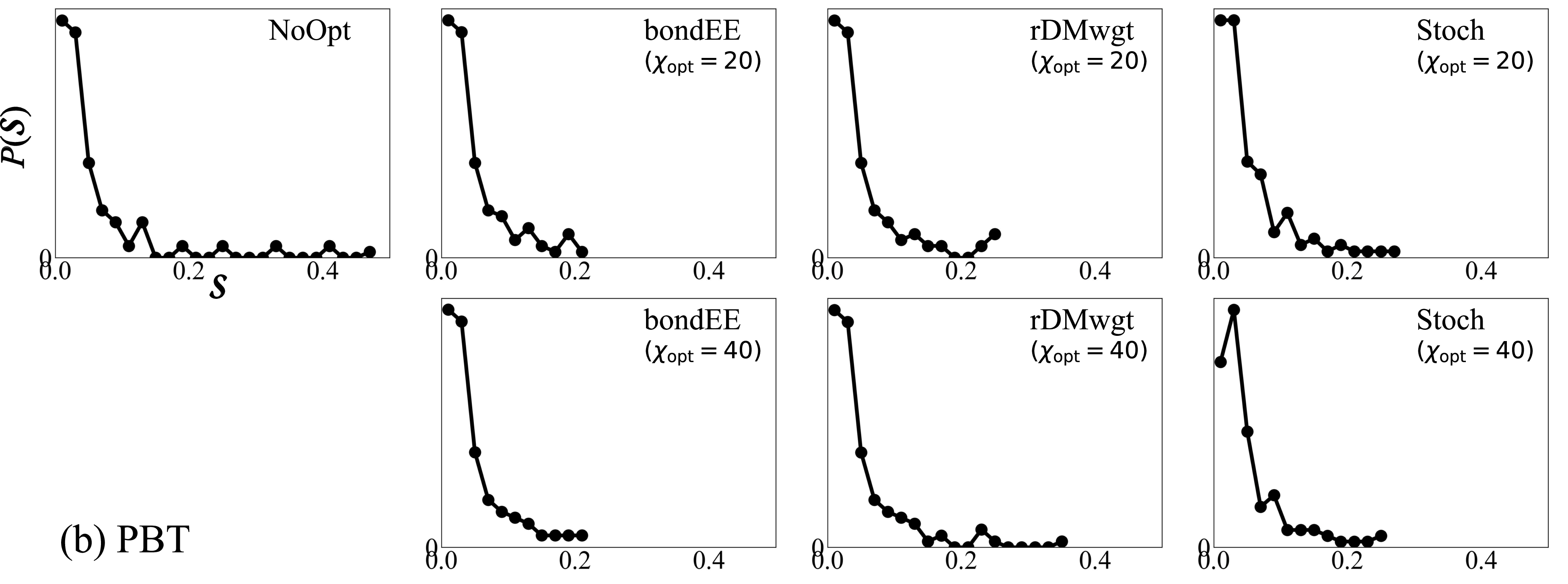}
\caption{
Histogram $P(\mathcal{S})$ of entanglement entropy $\mathcal{S}$ on auxiliary bonds in TTNs obtained by the calculations where the initial TTN is (a) MPN and (b) PBT for $J_{\rm R}=1$ and $N=128$.
``NoOpt" represents the calculation without structural optimization, whereas ``bondEE", ``rDMwgt", and ``Stoch" represent respectively the optimization using the bond EE, that using the truncated rDM weight, and the stochastic optimization.
Circles connected by solid lines represent the data at $\chi=80$.
The $\chi$ convergence has been confirmed by comparing the results at $\chi=80$ and $60$.
The size of bins is set to $\Delta \mathcal{S} = 0.02$.
For the stochastic optimization, the result of the run yielding the lowest variational energy at $\chi=80$ among the four runs is plotted.
}
\label{fig:EEhist_Rch}
\end{center}
\end{figure*}

The histograms of EEs on auxiliary bonds for $(J_{\rm R}, N)=(1, 128)$ are shown in Fig.\ \ref{fig:EEhist_Rch}.
The bin size is set to be $\Delta \mathcal{S} = 0.02$.
We find that the histogram obtained by the calculation without structural optimization for the initial MPN exhibits a unique stairway-like distribution.
The other histograms show a similar structure in which the frequency takes the largest value at a small EE and rapidly decreases as the EE increases.
The histograms for the other values of $J_{\rm R}$ and $N$ (not shown here) exhibit qualitatively the same behavior, while the scale of the horizontal axis (the values of EE) increases with $J_{\rm R}$.
It is thus suggested that in the Richardson model, the histograms of EE also have no strong correlation with the accuracy in the variational energy.

\begin{figure}
\begin{center}
\includegraphics[width = 80 mm]{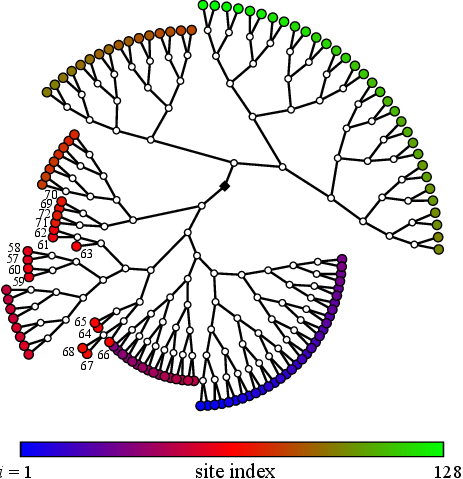}
\caption{
TTN structure obtained by the optimization using the truncated rDM weight for the initial PBT with $\chi_{\rm opt}=40$, which yielded the lowest variational energy at $\chi=80$, for $J_{\rm R}=1$ and $N=128$.
Open circles represent isometries and the solid diamond represents the singular-value matrix.
Bare spins are represented by circles with colors varying according to the site index $i$ as shown in the color bar.
Bare spins with the site index $57 \le i \le 72$ are indicated by numbers.
}
\label{fig:TTNoptimal_Rch}
\end{center}
\end{figure}

\begin{figure}
\begin{center}
\includegraphics[width = 80 mm]{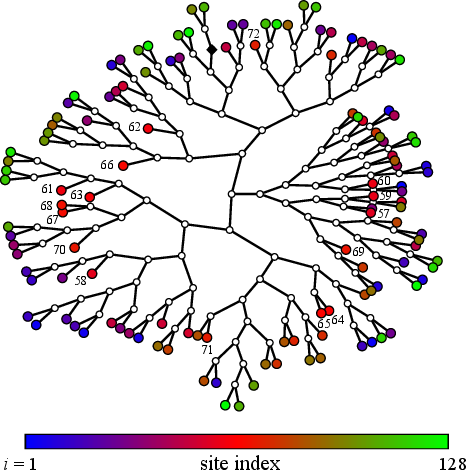}
\caption{
TTN structure obtained by a calculation of the stochastic optimization for the initial PBT with $\chi_{\rm opt}=40$, which yielded a high variational energy, for $J_{\rm R}=1$ and $N=128$.
Open circles represent isometries and the solid diamond represents the singular-value matrix.
Bare spins are represented by circles with colors varying according to the site index $i$ as shown in the color bar.
Bare spins with the site index $57 \le i \le 72$ are indicated by numbers.
}
\label{fig:TTNstcs_Rch}
\end{center}
\end{figure}

Finally, we examine the TTN structure obtained.
As discussed above, the optimal TTN for the Richardson model with Eq.\ (\ref{eq:hzi_Rch}) is expected to have a structure suitable for representing the entanglement among quantum-fluctuating spins around $i=N/2$.
Figure\ \ref{fig:TTNoptimal_Rch} presents the TTN structure obtained by the structural optimization using the truncated rDM weight for the initial PBT with $\chi_{\rm opt}=40$ for $J_{\rm R}=1$ and $N=128$, which yielded the lowest variational energy.
In this TTN, the spins around $i=N/2$ are located close to each other, indicating that a structure nearly the optimal one is realized.
(To be precise, as the spins at the sites $69 \le i \le 72$ are not far away but slightly apart from the clusters of the spins at $57 \le i \le 68$ in the TTN obtained, the TTN should not be the ``truly optimal", although it is close enough to the optimal to yield the lowest variational energy.)
Note that in the initial PBT of the present calculation, the spins around $i=N/2$ are separated into the two groups of $i \le N/2$ and $i \ge N/2+1$, which are located far apart;
Therefore, the initial PBT is not suitable for representing the significant entanglement across the two groups.
Our algorithm achieves a structural change from the initial PBT to the nearly optimal TTN.
We have also confirmed that the TTNs with the smallest variational energy obtained for other model parameters, $J_{\rm R}=1,4,16,64$ and $N=64$ have a similar structure in which the spins near $i=N/2$ are clustered close together, as in Fig.\ \ref{fig:TTNoptimal_Rch}.

The result that the stochastic optimization does not work well for the Richardson model also manifests itself in the TTN structure.
Figure\ \ref{fig:TTNstcs_Rch} shows the TTN structure obtained by the stochastic optimization for the initial PBT with $\chi_{\rm opt}=40$ for $J_{\rm R}=1$ and $N=128$.
The TTN of this structure yielded the relative error $\delta e = 5 \times 10^{-9}$ ($2 \times 10^{-8}$) for $\chi = 80$ ($60$), which seems quite small but $7\times 10^2$ ($2 \times 10^3$) times larger than the error by the optimal TTN.
In the TTN shown in Fig.\ \ref{fig:TTNstcs_Rch}, the spins around $i=N/2$ are scattered into various branches;
This TTN should fail in expressing the entanglement among those spins.
The reason why the stochastic optimization generated such a TTN may be understood as follows.
Roughly speaking, in the stochastic optimization, the TTN structure is randomized by probabilistic selection of local TTN structures at the early stage of the calculation where the effective temperature is high, which scatters the spins around $i=N/2$ into various branches.
Then, the TTN structure is optimized as the effective temperature is lowered.
However, in the randomized TTN, the spins near $i=N/2$ are surrounded by nearly classical-ordered spins at the sites away from $i=N/2$.
In the process of local reconnection of isometries, the position swap of the clusters of such classically ordered spins does not change the EE or the truncated rDM weight at the center bond very much.
As a result, the ``slope" of the cost function of the bond EE or the truncated rDM weight in the TTN-structural space becomes small.
In such a TTN, the structural optimization slows down and can be trapped even in a shallow local minimum.
This situation may not change even if $J_{\rm R}$ becomes larger, as long as the profile of $\langle S^z_i \rangle$ away from $i=N/2$ is flat and the effect of local reconnection of isometries on the cost function is small.

To confirm the above scenario, we have performed four runs of the non-stochastic structural optimization with $\chi_{\rm opt}=40$, starting from PBTs with random shuffling of bare spin sites.
The relative errors in the variational energy, $\delta e$, thus obtained are shown in Figs.\ \ref{fig:rltverr_m6080_Rch} and \ref{fig:rltverr_m80_Rch_J040-640} (by downward triangles).
The results indicate that the accuracy of the calculation is as poor as that of the stochastic optimization, supporting the above scenario that the stochastic optimization is roughly equivalent to the non-stochastic one starting from a randomized TTN.
These observations underscore the importance of appropriately selecting the initial TTN as well as the optimization scheme according to the system treated.

As for the non-stochastic optimizations, there are some cases where $\delta e^{(')}$ is relatively large, as seen in Figs.\ \ref{fig:rltverr_m6080_Rch} and \ref{fig:rltverr_m80_Rch_J040-640}.
We anticipate that this may also be understood by the scenario that once the structural optimization strays into the TTN structure with a small slope of the cost function, it is stuck there.

\section{Concluding Remarks}\label{sec:conc}

In summary, we have evaluated the performance of the TTN structural optimization algorithm for variational calculations of quantum many-body systems.
We have examined several types of optimization, which are specified by the initial TTN structure, the scheme to update the local connection of isometries, and the bond dimension at which the structural optimization is performed.
We applied them to the random XY-exchange model under random magnetic fields and the Richardson model and analyzed the variational energies and the EEs on the auxiliary bonds.
On the one hand, the results for the random XY model demonstrate that the TTN structural optimization algorithm works effectively to improve the accuracy of the calculation for the model.
In particular, the stochastic optimization, which probabilistically selects the local connection of isometries in the optimization process, yields the best results in the random average of the variational energy, regardless of the initial TTN structure.
On the other hand, the results for the Richardson model show that the structural optimization algorithm of an appropriately selected type can realize a lower variational energy than those obtained by the calculation without structural optimization.
Nevertheless, we have also found that, particularly in the calculations of the stochastic optimization, the TTN structure obtained by the optimization can get stuck in a structure that is not suitable for representing the lowest-energy state of interest, resulting in poor accuracy.

The TTN structural optimization algorithm we proposed searches for the optimal structure through local reconnections of the network.
Therefore, it is essential to devise a way to realize fast and steady convergence to the optimal structure.
We have tested the stochastic optimization scheme according to Eqs.\ (\ref{eq:stochas1}).
[See also Appendix\ \ref{app:Stcs2} for the test of another stochastic optimization using the probability of Eq.\ (\ref{eq:stochas2}).]
For the random XY model, the stochastic optimization succeeded in obtaining the lowest variational energies independent of the initial TTN structure, while the non-stochastic optimizations yielded results that are good but not the best and depend on the initial TTN.
In the random XY model, the landscape of the variational energy in the TTN-structure space is expected to be complex, and the optimization of the non-stochastic scheme is likely to be trapped in a local minimum, leading to poor accuracy.
The stochastic optimization should be effective in promoting the escape from such a local minimum.
In contrast, for the Richardson model, the stochastic optimization has yielded high variational energies, while the non-stochastic optimizations have achieved relatively good results depending on the initial TTN.
A possible explanation for this result is that once the TTN structure is randomized at the early stage of the stochastic optimization, the slope of the bond EE in the TTN structural space almost vanishes, and the optimization can be easily trapped in a shallow minimum.

The results of the present study indicate the importance of appropriately choosing the initial TTN structure and the scheme of selecting local connections of isometries in the structural optimization, depending on the model.
Generally speaking, for systems with randomness, it is naturally expected that the stochastic optimization is essential.
In contrast, for systems without randomness, the non-stochastic optimization should be the first choice since the stochastic scheme involves the risk that the structural optimization gets stuck in a randomized TTN structure.
Then, in either case, it must be effective to try the optimizations of various types as many as possible.
As observed in Sec\ \ref{sec:results}, the TTN structure that yields a good variational energy at a small $\chi$ usually produces a good energy also at a large $\chi$ (see Figs.\ \ref{fig:rltvge_XY} and \ref{fig:rltverr_Rch}).
Therefore, we can test several types of optimization calculations at a small $\chi$ where the computational cost is negligible, and then adopt the TTN structure with the best precision, which is expected to achieve the best precision also for a large $\chi$.
Another promising solution may be to employ a scheme based on the idea of the replica exchange method\cite{HukushimaN1996,RozadaAMK2019}, which has recently been applied to the structural optimization of tensor networks with loops\cite{WatanabeU2024}.

We have compared the accuracy of various types of structural optimizations at the same upper bound of the bond dimensions, $\chi$, focusing on the ability of TTNs constrained by the same $\chi$ to represent quantum states.
Here, it is also important to consider the computational costs to conduct the calculations.
In general, the computational cost of the calculation using a generic TTN is larger than that using MPN since isometries in MPN always have at least one physical bond with dimension $d$, which is often smaller than $\chi$, while isometries in TTN have no such constraint.
The memory required for storing isometries scales as $\mathcal{O}(N \chi^3)$ for generic (binary) TTN and $\mathcal{O}(N \chi^2 d)$ for MPN.
The computational costs to renormalize the spin and Hamiltonian operators are respectively $\mathcal{O}(N^2 \chi^4)$ and $\mathcal{O}(N \chi^5)$ for TTN whereas $\mathcal{O}(N^2 \chi^3 d)$ and $\mathcal{O}(N \chi^3 d^2)$ for MPN.
The cost to perform the update of isometries in the variational calculation depends on the algorithm;
For the two-tensor update algorithm, the cost scales as $\mathcal{O}(N N_{\rm int} \chi^5)$ for TTN and $\mathcal{O}(N N_{\rm int} \chi^3 d^2)$ for MPN, where $N_{\rm int}$ is a coefficient depending on the number of terms in the Hamiltonian.
For the single-tensor update, the cost is $\mathcal{O}(N N_{\rm int} \chi^4)$ for TTN and $\mathcal{O}(N N_{\rm int} \chi^3 d)$ for MPN.
Considering these differences, it may be a reasonable option to take advantage of the small cost of MPN, depending on the purpose and situation.
Alternatively, if one wishes to utilize the high capability of the optimal TTN for representing a quantum state, it should be effective to first perform the structural optimization of TTN at a small $\chi$ as mentioned above, and then improve isometries for a large $\chi$ by the single-tensor update algorithm, whose computational cost is relatively small.

The structural optimization based on the least-EE principle has been proven to be powerful not only for TTNs but also for MPNs\cite{LiRYS2022} and the tensor networks of entanglement renormalization.\cite{WatanabeU2024}
Meanwhile, it has been shown recently that for PEPS, the structural optimization based on the principle of selecting the network geometry that {\it maximizes} the bond EE is effective for improving the accuracy.\cite{PatraSO2025}
Although this {\it maximum}-EE principle for PEPS seems incompatible with the least-EE principle, these two are consistent with each other as follows.
In TTNs, each bond corresponds to a bipartition of the system that occurs when the bond is cut.
The least-EE principle minimizes the appearance of bonds corresponding to a bipartition with a large EE.
However, when a physical spin is strongly entangled with other spins, the TTN must represent the large entanglement precisely.
Consequently, in a TTN optimized according to the least-EE principle, the strongly entangled spins are connected by as few isometries as possible in the lower layers (regions near the boundary) of the network.
In such a TTN, the bonds with large EE that are inevitable for representing the target state accurately are confined to small regions in lower layers, thereby minimizing the bond EEs in the whole TTN.
In fact, our calculation of the structural optimization for the Richardson model has yielded the TTN structure in which the spins entangled strongly are located close to each other.
Similar TTNs have also been obtained for the hierarchical chain\cite{HikiharaUOHN2023a} and the Rainbow chain.\cite{HikiharaUOHN2023b}
On the other hand, the optimization principle for PEPS that maximizes the bond EEs yields the network structure in which the strongly entangled spins are connected by a small number of bonds (directly by a single bond, if possible), making the distance between those strongly entangled spins in the PEPS network as small as possible.
It is thereby understood that the {\it maximum}-EE principle for PEPS realizes the situation occurring in the lower layer of the TTN obtained by the least-EE principle.
We note that a similar strategy to connect the strongly entangled spins at the lower layer of the TTN has also been employed in the entanglement-based tSDRG.\cite{SekiHO2021}

While we have focused on the performance of the TTN structural optimization algorithm in the variational calculation of quantum many-body systems, it has many other potential applications.
A promising application is an efficient representation of a given high-rank tensor in the TTN format.
In the algorithm we proposed, each step involves the diagonalization of the effective Hamiltonian expanded in the truncated Hilbert space of the central area to obtain the ground-state wave function (process 3 of Table\ \ref{tab:algorithm}).
By replacing this process with the one to obtain a wave function that has a maximum fidelity to the given high-rank tensor, one can optimize only the TTN structure without changing the TTN state itself.\cite{WatanabeMHU2025}
Such an application is expected to be useful for machine learning\cite{StoudenmireS2016,ChengWXZ2019} and quantics tensor-train approach\cite{ShinaokaWMNSWK2023,RitterNWDSW2024,Fernandez2024}, to which tensor networks have been applied actively in recent years.
An application of a similar algorithm using bond mutual information as a cost function to the Born machine has been explored very recently.\cite{Harada2024OK2024}

\begin{acknowledgments}
This work is partially supported by KAKENHI Grant Numbers JP21H04446, JP22H01171, JP24K06881, and a Grant-in-Aid for Transformative Research Areas titled ``The Natural Laws of Extreme Universe-A New Paradigm for Spacetime and Matter from Quantum Information" (KAKENHI Grants No. JP21H05182, No. JP21H05191) from JSPS of Japan.
We also acknowledge support from MEXT Q-LEAP Grant No. JPMXS0120319794, JST COI-NEXT No. JPMJPF2014, and JST-CREST No. JPMJCR24I1.
H.U. and T. N. were supported by the COEresearch grant in computational science from HyogoPrefecture and Kobe City through Foundation for Computational Science.
\end{acknowledgments}

The data that support the findings of this article are openly available in \cite{data_availability}.

\appendix
\section{Preparation of initial TTN}\label{app:initialTTN}

In our calculation, isometries in the initial TTN are prepared as follows.
For example, let us consider the MPN shown in Fig.\ \ref{fig:TTNs} (a).
First, we focus on the subsystem consisting of the leftmost two spins at the sites $i=1$ and $2$ in the MPN and construct its block Hamiltonian.
Then, we generate the leftmost isometry using the $\chi$ eigenvectors of the block Hamiltonian with the $\chi$-lowest eigenenergies.
If the number of eigenvectors of the block Hamiltonian does not exceed $\chi$, all eigenvectors are adopted in the isometry.
Next, we focus on the subsystem consisting of the leftmost two-site block and the neighboring spin at the site $i=3$, construct and diagonalize the block Hamiltonian of the subsystem, and generate the new isometry using the $\chi$-lowest-energy eigenvectors.
We iterate these procedures from the left edge to the center of the MPN, where the canonical center is located, and perform the same operation from the right edge to the center to eventually prepare the initial MPN.
(To be precise, we prepare only the isometries in the initial TTN and do not generate the singular-value matrix, which is not necessary for the start of our calculation.)
The other initial TTNs are also prepared in the same way, with the only difference that we change the order of merging the spin blocks.
In the case of PBT, the spin blocks are merged from the boundary of PBT, where the bare spins are located, to the ``top" of PBT, where the canonical center is, so that PBT shown in Fig.\ \ref{fig:TTNs} (b) is constructed.
In the case of the tSDRG, in each step of the preparation procedure, we construct the interaction Hamiltonians of all the pairs of spin blocks, diagonalize them, and calculate the maximum level spacing $\Delta_{\rm max}^{\rm I}$ of their energy spectrum.
Then, among all the spin-block pairs, we select the one with the largest $\Delta_{\rm max}^{\rm I}$ to be merged at the step.
See Ref.\ \cite{SekiHO2020} for the details.

In general, in our algorithm, the structure of the initial TTN can be arbitrary.
The closer the structure of the initial TTN is to the optimal one, the faster the convergence of the calculation is.
As for the initial isometries, any isometries can be used as long as they satisfy the orthonormal condition Eq.\ (\ref{eq:orthonormal}).
Preparing the isometries using random vectors satisfying the condition may be a reasonable choice.
It is also possible to use a product state, a TTN with $\chi=1$, as the initial TTN.
However, in that case, it is necessary to take measures against the problem of the appearance of an isometry whose three legs have the dimension unity, mentioned in Sec.\ \ref{sec:algo}, since the TTN with $\chi=1$ of a generic structure inevitably contains such an isometry and the calculation will be stuck.
A possible measure to avoid the problem is employing an MPN as the initial TTN so that the central area has at least two physical bonds, whose dimension is usually larger than unity, and increasing the dimension of auxiliary bonds sufficiently before starting the structural optimization.

The problem of the isometry having three legs of dimension unity may also occur accidentally during the TTN optimization calculation.
In the present work, we encountered the problem in several cases of the calculations where the initial TTN was PBT for the Richardson model with $(J_{\rm R}, N)=(64,64)$, if we did not take the exceptional treatment mentioned below.
The problem occurred when the initial PBT included isometries whose outgoing bond was composed only of the bases having the same magnetization.
Since the states of the spin block represented by such an isometry can take only the single magnetization in the truncated Hilbert space, the XY-exchange interactions between the block and the rest of the system become zero, and the block is decoupled.
Then, if two or more such blocks are connected to the central area,
the rank of the ground-state wave function in one (or more) singular-value decomposition becomes unity, generating a bond with dimension $\chi=1$.
The cumulation of such a process may result in the emergence of the isometry with all three legs having the dimension $\chi=1$.

In order to avoid such a situation, for the Richardson model, we adopted an exceptional treatment in the procedure to prepare the initial PBT;
Namely, if all of the $\chi$-lowest-energy eigenstates of the block Hamiltonian had the same magnetization, we discarded the state with the highest energy (or the highest-energy multiplet if the states were degenerate) and incorporated the lowest-energy state in the subspaces with the one higher and one lower magnetization to construct the isometry.
We succeeded in avoiding the problem with this treatment.

\section{Details of calculations}\label{app:Detail_Calcu}

In our TTN calculations, we computed physical quantities such as variational energies as a function of the upper bound of the bond dimension, $\chi$, by increasing $\chi$.
The calculations for the random XY model and the Richardson model were performed in the following manner.

{\it random XY model}:
For the calculation without structural optimization, we performed the variational calculation to optimize the isometries by increasing $\chi$ as $\chi = 16 (100), 24 (20), 32 (20), 40 (20)$, where the figures in the parentheses are the {\it maximum} number of sweeps iterated at each $\chi$.
If the maximum number of sweeps was reached or the variational energy converged within the range of $10^{-5}$, we finished the calculation at the value of $\chi$ and proceeded to the next $\chi$.
We then adopted the lowest variational energy obtained during the sweeps at each $\chi$ as the variational energy for the value of $\chi$.

For the calculations with the structural optimization, we first performed the calculations with the structural optimization procedure at $\chi=\chi_{\rm opt}$;
Two cases of calculations with $\chi_{\rm opt}=16$ and $32$ were carried out.
We iterated the structural optimization calculations until the TTN structure and the variational energy converged, and then adopted the TTN structure obtained as the optimized one in the following calculation.
If the calculation did not converge within $80$ sweeps, the TTN structure obtained after the last sweep was used as the optimized one.
After the optimized TTN structure was determined, we performed the variational calculation to optimize the isometries by increasing $\chi$ as $\chi = 16, 24, 32, 40$ for $\chi_{\rm opt}=16$ and $\chi=32, 40$ for $\chi_{\rm opt}=32$.
The calculation at each $\chi$ was continued until the variational energy converged or the number of sweeps reached $20$.

{\it Richardson model}: 
The calculations for the Richardson model were performed in a manner similar to those for the random XY model, with slight differences (see below).

For the calculation without structural optimization, we performed the calculation to optimize the isometries by increasing $\chi$ as $\chi = 20 (100), 40 (20), 60 (20), 80 (20)$, where the figures in the parentheses denote the number of sweeps iterated at each $\chi$.
Here, differently from the case of the random XY model, we did not monitor the convergence of the variational energy but instead performed the calculation of the stated number of sweeps.
We then adopted the lowest variational energy obtained during the sweeps as the variational energy at the value of $\chi$.
For the calculations with the structural optimization, we performed the calculations with the structural optimization procedure at $\chi=\chi_{\rm opt}=20$ or $40$ until the convergence of the TTN structure and the variational energy (within the range of $10^{-6}$) was achieved or the number of sweeps reached $80$.
After that, we performed the variational calculation to optimize the isometries by increasing $\chi$ as $\chi = 20, 40, 60, 80$ for $\chi_{\rm opt}=20$ and $\chi=40, 60, 80$ for $\chi_{\rm opt}=40$.
We iterated the calculations of 20 sweeps for each $\chi$ to obtain the variational energy as a function of $\chi$.

\section{Results of stochastic calculation using the probability $P^{(2)}$ in Eq.\ (\ref{eq:stochas2})}\label{app:Stcs2}

\begin{figure*}
\begin{center}
\includegraphics[width = 48 mm]{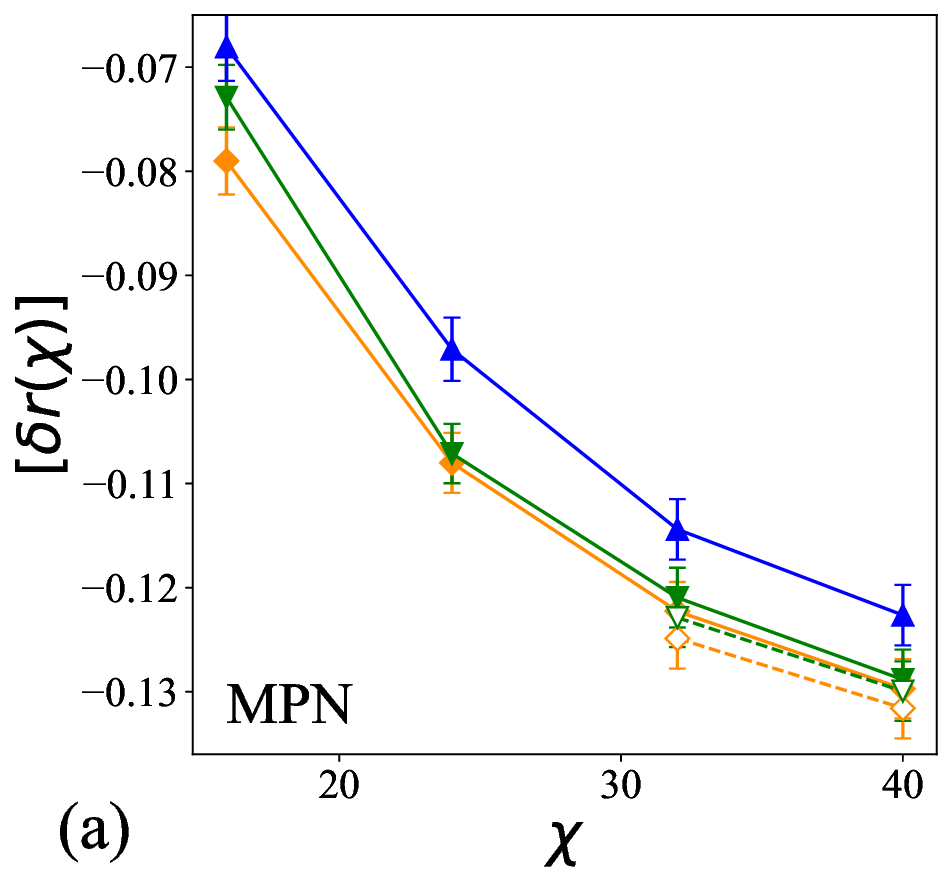}
\includegraphics[width = 48 mm]{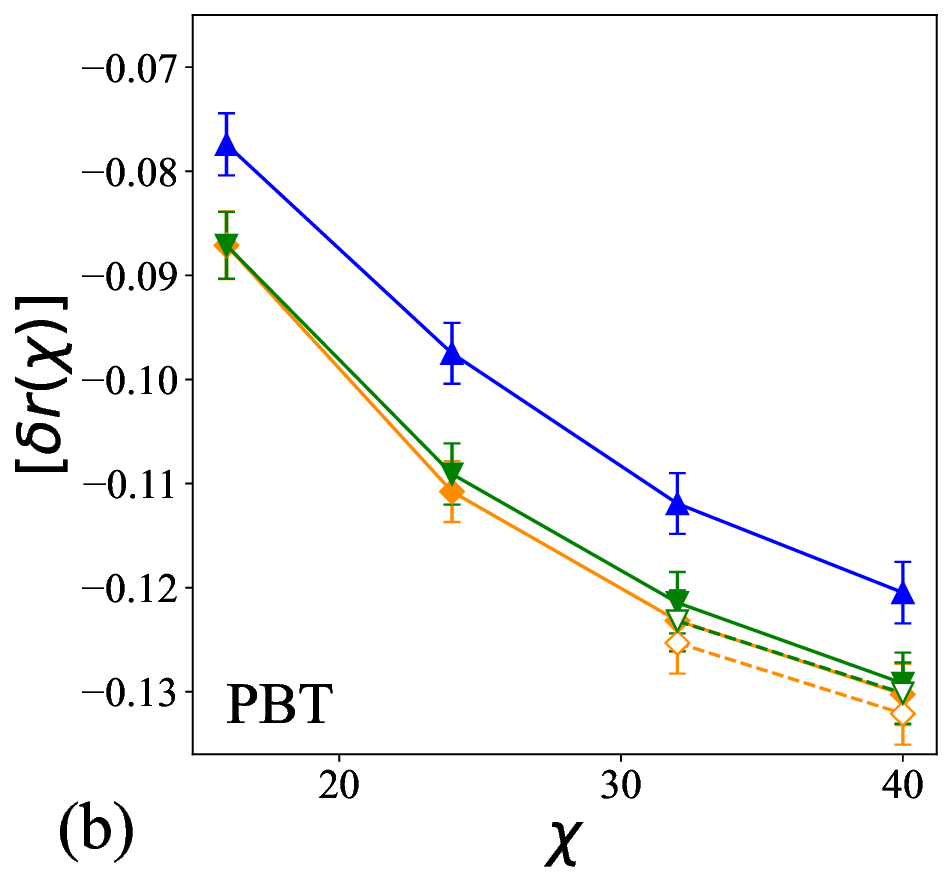}
\includegraphics[width = 74 mm]{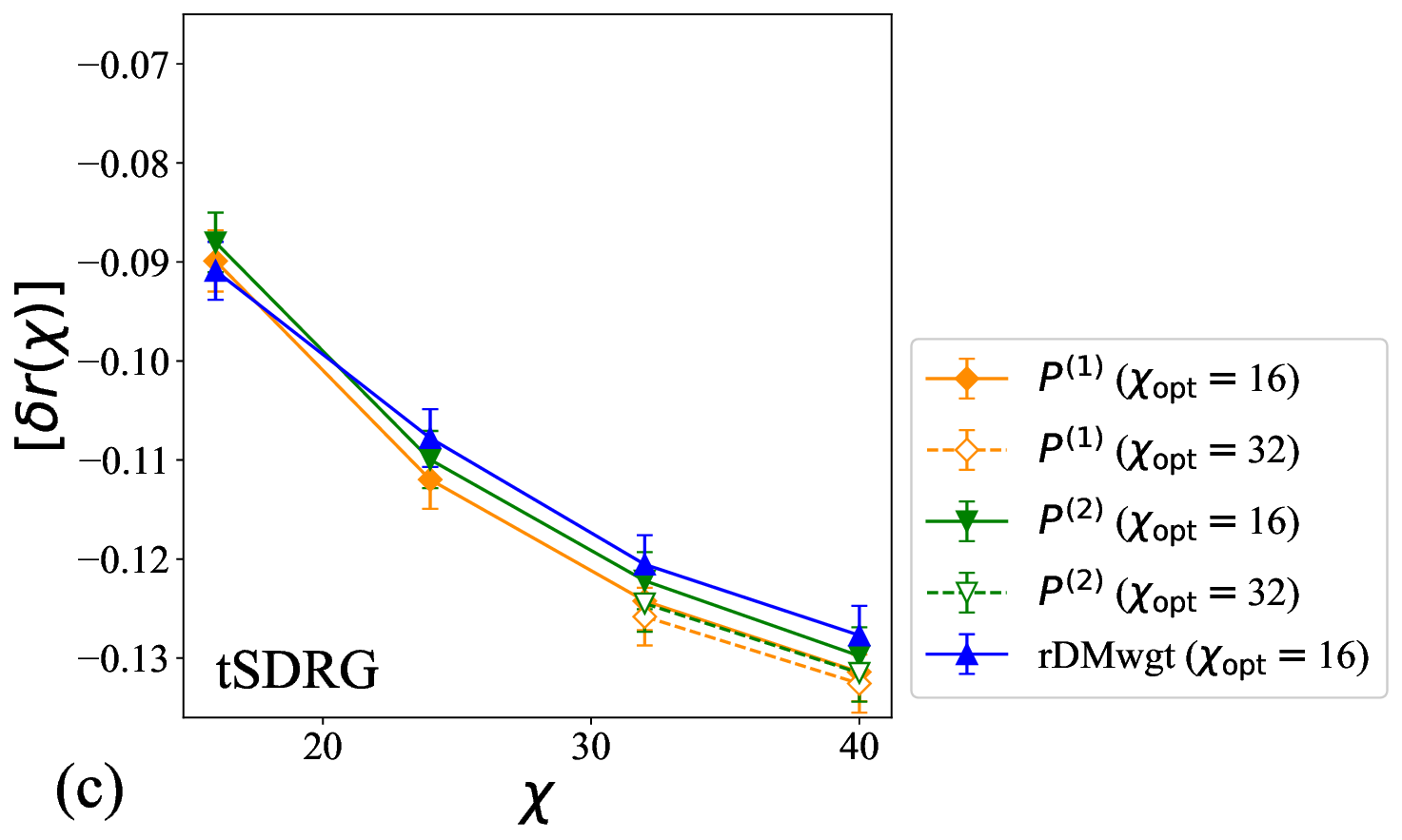}
\caption{
Random averages of the relative reduction in the variational energy, $[\delta r(\chi)]$, for the random XY model obtained by the calculations where the initial TTN is (a) MPN, (b) PBT, and (c) tSDRG-TTN.
``$P^{(1)}$" and ``$P^{(2)}$" represent respectively the stochastic optimizations according to the probability $P^{(1)}$ [Eq.\ (\ref{eq:stochas1})] and $P^{(2)}$ [Eq.\ (\ref{eq:stochas2})].
The results for the optimization using the truncated rDM weight (``rDMwgt") are also presented for comparison.
}
\label{fig:stochas2-XYrand}
\end{center}
\end{figure*}

\begin{figure*}
\begin{center}
\includegraphics[width = 48 mm]{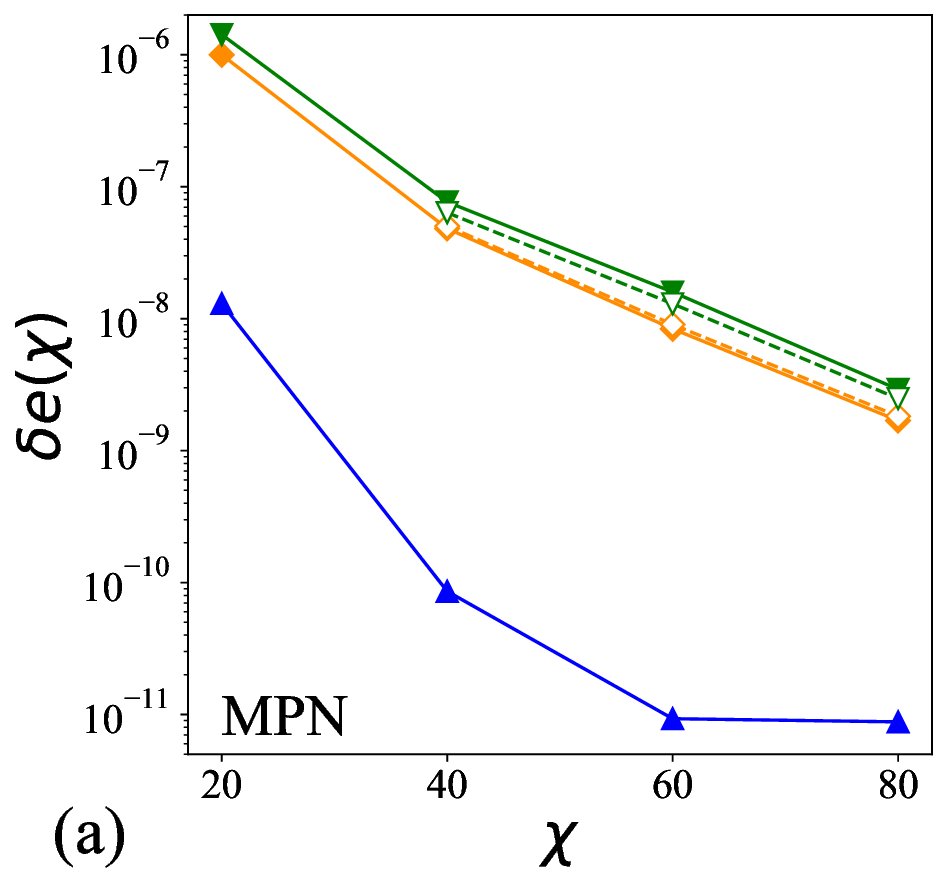}
\includegraphics[width = 74 mm]{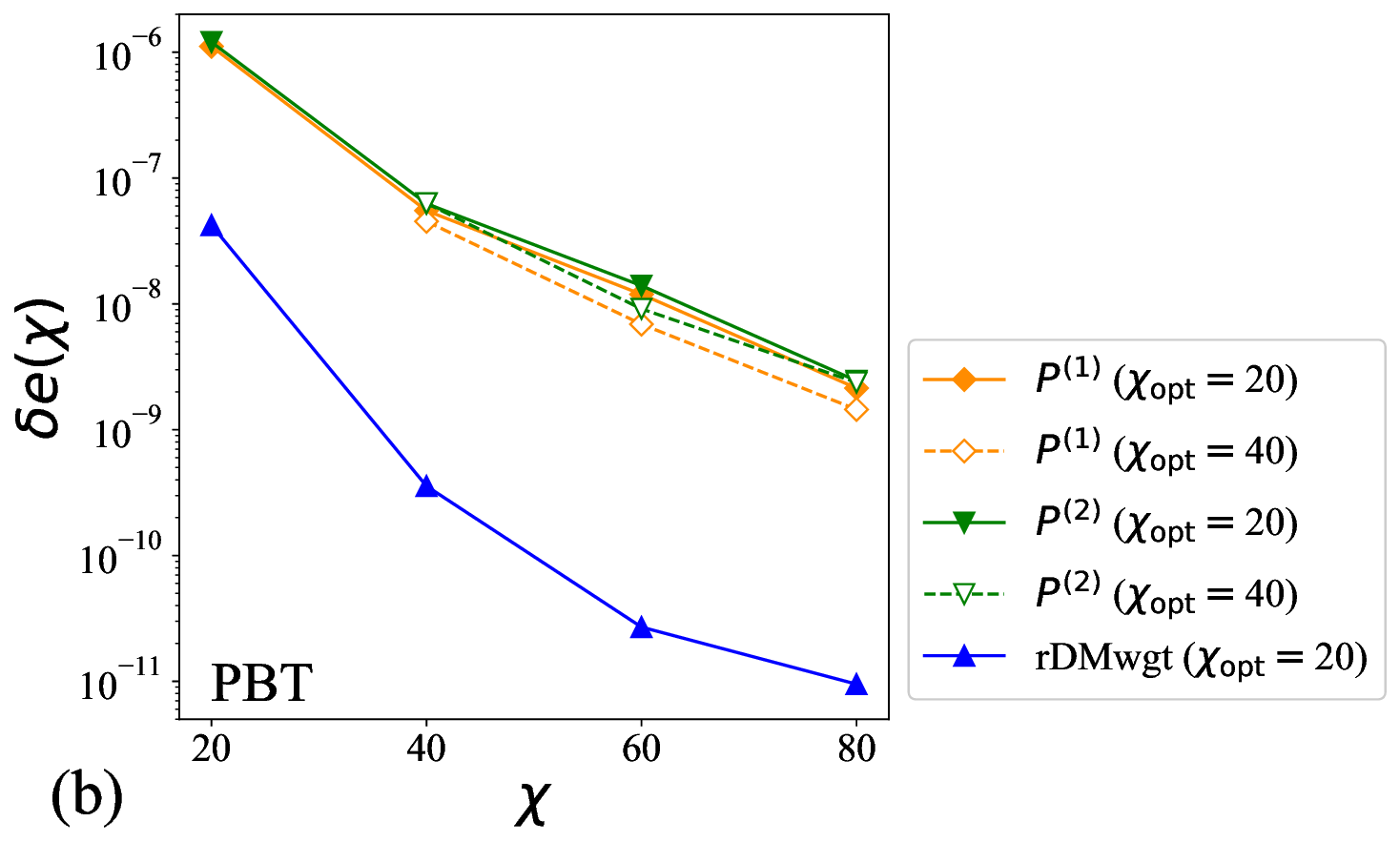}
\caption{
Relative error in the variational energy, $\delta e(\chi)$, for the Richardson model obtained by the calculations where the initial TTN is (a) MPN and (b) PBT.
The model parameters are $J_{\rm R}=1$ and $N=128$.
``$P^{(1)}$" and ``$P^{(2)}$" represent respectively the stochastic optimizations according to the probability $P^{(1)}$ [Eq.\ (\ref{eq:stochas1})] and $P^{(2)}$ [Eq.\ (\ref{eq:stochas2})].
Among the four runs, the result of the run yielding the lowest variational energy at $\chi=80$ is plotted.
The results for the optimization using the truncated rDM weight (``rDMwgt") are also presented for comparison.
}
\label{fig:stochas2-Rch}
\end{center}
\end{figure*}

In the stochastic optimization of the TTN structure discussed in the main text, we have performed the probabilistic selection of the local connection of the isometries according to the probability $P^{(1)}$ in Eq.\ (\ref{eq:stochas1}).
As mentioned in Secs.\ \ref{sec:algo} and \ref{subsec:scheme}, it is also possible to adopt the probability $P^{(2)}$ in Eq.\ (\ref{eq:stochas2}) for the probabilistic selection, instead of $P^{(1)}$.
We have also performed the calculations of the stochastic optimization with $P^{(2)}$ for all cases where the calculations with $P^{(1)}$ were carried out, for both the random XY model and the Richardson model.
We have then compared the performance of those calculations.

Figures\ \ref{fig:stochas2-XYrand} and \ref{fig:stochas2-Rch} show the results of the relative reduction $[\delta r(\chi)]$ and the relative error $\delta e(\chi)$ of the variational energy for the random XY model and the Richardson model, respectively, obtained by the stochastic optimizations with $P^{(1)}$ and $P^{(2)}$.
(Note that the data of the calculations using $P^{(1)}$ are the same as those shown in Figs.\ \ref{fig:rltvge_XY} and \ref{fig:rltverr_Rch} in the main text.)
We find that both optimization schemes achieve almost the same accuracy for those models:
while the results of the optimization using $P^{(1)}$ may be slightly more accurate than those using $P^{(2)}$, the differences are minor compared to the differences from the results of the other structural optimization schemes.
In addition, we have analyzed the average values, maximum values, and the histograms of EEs on the auxiliary bonds, and confirmed that the stochastic optimizations with $P^{(1)}$ and $P^{(2)}$ yield essentially the same results.
Therefore, we conclude that the stochastic optimizations using $P^{(1)}$ and $P^{(2)}$ exhibit nearly equivalent performance, at least for the models treated in the present study.
In practice, it may be sufficient to perform the calculation with $P^{(1)}$, and there is no particular need to execute the additional calculation with $P^{(2)}$.

\section{Exact solution of Richardson model}\label{app:Exact_Rch}

The Richardson model given by Eq.\ (\ref{eq:Ham_Rch}) is exactly solvable.
We calculated the exact ground-state energy of the model using the equations and method summarized below.
Here, we assume that the magnetic fields $\{ h_i \}$ are non-degenerate.
We refer the readers to Ref.\ \cite{DelftR2001} for more details.

First, we rewrite the spin Hamiltonian $\mathcal{H}_{\rm R}$ in Eq.\ (\ref{eq:Ham_Rch}) in terms of hardcore bosons.
The spin-1/2 operators can be represented using the hardcore boson operators as
\begin{eqnarray}
S^+_i = b_i,~~~
S^-_i = b_i^\dagger,~~~
S^z_i = \frac{1}{2} - b_i^\dagger b_i,
\label{eq:Spin_to_boson}
\end{eqnarray}
where $S^\pm_i = S^x_i \pm i S^y_i$, and $b_i$ is the hardcore-boson operator at the $i$ th site obeying $\left[ b_i, b_{i'}^\dagger \right] = \left( 1 - 2 b_i^\dagger b_i \right) \delta_{i i'}$ and $\left( b_i^\dagger \right)^2 = 0$.
Substituting the expressions of Eq.\ (\ref{eq:Spin_to_boson}) into the spin Hamiltonian $\mathcal{H}_{\rm R}$, one can map the spin model into the interacting hardcore-boson model $\tilde{\mathcal{H}}_{\rm R}$ as
\begin{eqnarray}
\mathcal{H}_{\rm R} &=& \tilde{\mathcal{H}}_{\rm R} - \frac{1}{2} \sum_{i=1}^N h_i + g N_{\rm b},
\label{eq:spinHam_to_bosonHam} \\
\tilde{\mathcal{H}}_{\rm R} &=& \sum_{i=1}^N \sum_{i'=1}^N \left( h_i \delta_{ii'} - g \right) b^\dagger_i b_{i'}.
\label{eq:Ham_Rch_boson}
\end{eqnarray}
The coupling constant $g$ and the number of hardcore bosons, $N_{\rm b}$, are respectively given by
\begin{eqnarray}
g &=& - \frac{J_{\rm R}}{2 N},
\label{eq:g_to_J} \\
N_{\rm b} &=& \frac{N}{2} -M,
\label{eq:Num_boson}
\end{eqnarray}
where $M = \sum_{i=1}^N S^z_i$ is the total magnetization.
We consider the case of even $N_{\rm b}$ in the following.

The eigenstates of the Hamiltonian $\tilde{\mathcal{H}}_{\rm R}$ [Eq.\ (\ref{eq:Ham_Rch_boson})] can be written down explicitly as
\begin{eqnarray}
|\tilde{\Psi}_{\rm R} \rangle &=& \prod_{n = 1}^{N_{\rm b}} B^\dagger_n |0\rangle,
\label{eq:eigenstate_Rch} \\
B^\dagger_n &=& \sum_{i=1}^N \frac{b^\dagger_i}{h_i - E_n},
\label{eq:B-operator}
\end{eqnarray}
where $| 0 \rangle$ is the vacuum of the bosons $\{ b_i \}$.
The state $|\tilde{\Psi}_{\rm R} \rangle$ in Eq.\ (\ref{eq:eigenstate_Rch}) is an exact eigenstate of the Hamiltonian $\tilde{\mathcal{H}}_{\rm R}$ if the parameters $\{ E_n \}$ satisfy the coupled equations,
\begin{eqnarray}
1 - \sum_{i=1}^N \frac{g}{h_i - E_n} + \sum_{n' = 1 (\ne n)}^{N_{\rm b}} \frac{2g}{E_{n'} - E_n} = 0,
\label{eq:coupled_eqs}
\end{eqnarray}
for $n = 1, ..., N_{\rm b}$.
The eigenenergy is obtained as
\begin{eqnarray}
\mathcal{E} = \sum_{n=1}^{N_{\rm b}} E_n.
\label{eq:eigenenergy_Rch}
\end{eqnarray}

The solution $\{ E_n \}$ of the coupled equations (\ref{eq:coupled_eqs}) consists of real numbers or complex conjugate pairs.
We thus express $\{ E_n \}$ as
\begin{eqnarray}
E_{2p-1} = x_p - i y_p,~~~
E_{2p}   = x_p + i y_p,
\label{eq:E_to_xy}
\end{eqnarray}
for $p = 1, ..., N_{\rm b}/2$, where $x_p$ are real and $y_p$ are real or pure imaginary.
The coupled equations (\ref{eq:coupled_eqs}) are rewritten as
\begin{eqnarray}
&&1 - g \sum_{i=1}^N \frac{h_i - x_p}{(h_i - x_p)^2 + y_p^2}
\nonumber \\
&&~~+ 4g \sum_{p'=1 (\ne p)}^{N_{\rm b}/2} \frac{(x_{p'}-x_p)\left[ (x_{p'}-x_p)^2+y_p^2+y_{p'}^2\right]}{\left[(x_{p'}-x_p)^2+y_p^2+y_{p'}^2\right]^2-4y_p^2y_{p'}^2} = 0,
\nonumber \\
\label{eq:coupled_eqs_xy1} \\
&&1 - \sum_{i=1}^N \frac{y_p^2}{(h_i - x_p)^2 + y_p^2}
\nonumber \\
&&~~+ 4 \sum_{p'=1 (\ne p)}^{N_{\rm b}/2} \frac{y_p^2\left[ (x_{p'}-x_p)^2+y_p^2+y_{p'}^2\right] - 2y_p^2y_{p'}^2}{\left[(x_{p'}-x_p)^2+y_p^2+y_{p'}^2\right]^2-4y_p^2y_{p'}^2} = 0.
\nonumber \\
\label{eq:coupled_eqs_xy2}
\end{eqnarray}
Note that in these equations, $y_p$ appears only in the form of $y_p^2$.
Therefore, we can treat $\{ x_p \}$ and $\{ y_p^2 \}$ as real parameters and solve the coupled equations (\ref{eq:coupled_eqs_xy1}) and (\ref{eq:coupled_eqs_xy2}) numerically using the Newton-Raphson method\cite{NumRecipes}.

It should be noted that the above coupled equations may yield not only the ground state but also an arbitrary eigenstate of the Richardson model.
We obtained the ground-state solution in the following manner.
First, the ground-state solution at $g=0$ is trivially obtained as
\begin{eqnarray}
E_n(g=0) = h_n,~~(n=1,...,N_{\rm b})
\end{eqnarray}
or equivalently,
\begin{eqnarray}
x_p(g=0) &=& \frac{h_{2p}+h_{2p-1}}{2},
\\
y_p^2(g=0) &=& - \left( \frac{h_{2p}-h_{2p-1}}{2}\right)^2.
\end{eqnarray}
Here, note that the field $h_i$ is arranged in ascending order.
Furthermore, for $g = - \delta g$ with a small $\delta g > 0$, the approximate ground-state solution up to the first order of $\delta g$ is obtained as
\begin{eqnarray}
\bar{x}_p(g=-\delta g) &=& \frac{h_{2p}+h_{2p-1}}{2} + \delta g,
\\
\bar{y}_p^2(g=-\delta g) &=& - \left( \frac{h_{2p}-h_{2p-1}}{2}\right)^2.
\end{eqnarray}
We adopted these values as initial guesses and input them into the Newton-Raphson method to obtain the ground-state solution for $g = - \delta g$.
Afterwards, we prepared the initial guesses for $g = - n ~ \delta g$ ($n = 2, 3, 4, ...$) as
\begin{eqnarray}
\bar{x}_p(g=-n ~ \delta g) &=& 2 x_p(g=-(n-1) \delta g) 
\nonumber \\
&&- x_p(g=-(n-2)\delta g),
\\
\bar{y}_p^2(g=-n ~ \delta g) &=& 2 y_p^2(g=-(n-1) \delta g) 
\nonumber \\
&&- y_p^2(g=-(n-2) \delta g),
\end{eqnarray}
and input them into the Newton-Raphson method.
In such a manner, we obtained the ground-state solutions as a function of $g$ by decreasing $g$ [increasing $J_{\rm R}$ in Eq.\ (\ref{eq:g_to_J})] adiabatically.
The obtained ground-state energies for the model parameters discussed in Sec.\ \ref{subsec:Rch} are presented in Table\ \ref{tab:exact_energy_Rch}.

\begin{table}
\caption{
Exact ground-state energy of the Richardson model [Eq.\ (\ref{eq:Ham_Rch})] with the magnetic field term of Eq.\ (\ref{eq:hzi_Rch}) for the model parameters $(J_{\rm R}, N) = (1.0, 64), (4.0, 64), (16.0, 64), (64.0, 64)$, and $(1.0, 128)$.
}
\label{tab:exact_energy_Rch}
\begin{center}
\begin{tabular}{rrc}
\hline
\hline
 $J_{\rm R}$ & $N$ & $E_{\rm ex}$ \\
\hline
 1.0 &  64 &  -8.20139294258928 \\
 4.0 &  64 &  -8.64996248818904(1) \\
16.0 &  64 &  -11.00530807225508 \\
64.0 &  64 &  -21.76617777856154(1) \\
 1.0 & 128 &  -16.20027338853789 \\
\hline
\hline
\end{tabular}
\end{center}
\end{table}


\begin{thebibliography}{49}%
\makeatletter
\providecommand \@ifxundefined [1]{%
 \@ifx{#1\undefined}
}%
\providecommand \@ifnum [1]{%
 \ifnum #1\expandafter \@firstoftwo
 \else \expandafter \@secondoftwo
 \fi
}%
\providecommand \@ifx [1]{%
 \ifx #1\expandafter \@firstoftwo
 \else \expandafter \@secondoftwo
 \fi
}%
\providecommand \natexlab [1]{#1}%
\providecommand \enquote  [1]{``#1''}%
\providecommand \bibnamefont  [1]{#1}%
\providecommand \bibfnamefont [1]{#1}%
\providecommand \citenamefont [1]{#1}%
\providecommand \href@noop [0]{\@secondoftwo}%
\providecommand \href [0]{\begingroup \@sanitize@url \@href}%
\providecommand \@href[1]{\@@startlink{#1}\@@href}%
\providecommand \@@href[1]{\endgroup#1\@@endlink}%
\providecommand \@sanitize@url [0]{\catcode `\\12\catcode `\$12\catcode
  `\&12\catcode `\#12\catcode `\^12\catcode `\_12\catcode `\%12\relax}%
\providecommand \@@startlink[1]{}%
\providecommand \@@endlink[0]{}%
\providecommand \url  [0]{\begingroup\@sanitize@url \@url }%
\providecommand \@url [1]{\endgroup\@href {#1}{\urlprefix }}%
\providecommand \urlprefix  [0]{URL }%
\providecommand \Eprint [0]{\href }%
\providecommand \doibase [0]{https://doi.org/}%
\providecommand \selectlanguage [0]{\@gobble}%
\providecommand \bibinfo  [0]{\@secondoftwo}%
\providecommand \bibfield  [0]{\@secondoftwo}%
\providecommand \translation [1]{[#1]}%
\providecommand \BibitemOpen [0]{}%
\providecommand \bibitemStop [0]{}%
\providecommand \bibitemNoStop [0]{.\EOS\space}%
\providecommand \EOS [0]{\spacefactor3000\relax}%
\providecommand \BibitemShut  [1]{\csname bibitem#1\endcsname}%
\let\auto@bib@innerbib\@empty
\bibitem [{\citenamefont {White}(1992)}]{White1992}%
  \BibitemOpen
  \bibfield  {author} {\bibinfo {author} {\bibfnamefont {S.~R.}\ \bibnamefont
  {White}},\ }\bibfield  {title} {\bibinfo {title} {Density matrix formulation
  for quantum renormalization groups},\ }\href
  {https://doi.org/10.1103/PhysRevLett.69.2863} {\bibfield  {journal} {\bibinfo
   {journal} {Phys. Rev. Lett.}\ }\textbf {\bibinfo {volume} {69}},\ \bibinfo
  {pages} {2863} (\bibinfo {year} {1992})}\BibitemShut {NoStop}%
\bibitem [{\citenamefont {White}(1993)}]{White1993}%
  \BibitemOpen
  \bibfield  {author} {\bibinfo {author} {\bibfnamefont {S.~R.}\ \bibnamefont
  {White}},\ }\bibfield  {title} {\bibinfo {title} {Density-matrix algorithms
  for quantum renormalization groups},\ }\href
  {https://doi.org/10.1103/PhysRevB.48.10345} {\bibfield  {journal} {\bibinfo
  {journal} {Phys. Rev. B}\ }\textbf {\bibinfo {volume} {48}},\ \bibinfo
  {pages} {10345} (\bibinfo {year} {1993})}\BibitemShut {NoStop}%
\bibitem [{\citenamefont {Or{\'u}s}(2019)}]{Orus2019}%
  \BibitemOpen
  \bibfield  {author} {\bibinfo {author} {\bibfnamefont {R.}~\bibnamefont
  {Or{\'u}s}},\ }\bibfield  {title} {\bibinfo {title} {Tensor networks for
  complex quantum systems},\ }\href {https://doi.org/10.1038/s42254-019-0086-7}
  {\bibfield  {journal} {\bibinfo  {journal} {Nature Reviews Physics}\ }\textbf
  {\bibinfo {volume} {1}},\ \bibinfo {pages} {538} (\bibinfo {year}
  {2019})}\BibitemShut {NoStop}%
\bibitem [{\citenamefont {Okunishi}\ \emph {et~al.}(2022)\citenamefont
  {Okunishi}, \citenamefont {Nishino},\ and\ \citenamefont
  {Ueda}}]{OkunishiNU2022}%
  \BibitemOpen
  \bibfield  {author} {\bibinfo {author} {\bibfnamefont {K.}~\bibnamefont
  {Okunishi}}, \bibinfo {author} {\bibfnamefont {T.}~\bibnamefont {Nishino}},\
  and\ \bibinfo {author} {\bibfnamefont {H.}~\bibnamefont {Ueda}},\ }\bibfield
  {title} {\bibinfo {title} {Developments in the tensor network - from
  statistical mechanics to quantum entanglement},\ }\href
  {https://doi.org/10.7566/JPSJ.91.062001} {\bibfield  {journal} {\bibinfo
  {journal} {Journal of the Physical Society of Japan}\ }\textbf {\bibinfo
  {volume} {91}},\ \bibinfo {pages} {062001} (\bibinfo {year}
  {2022})}\BibitemShut {NoStop}%
\bibitem [{\citenamefont {Larsson}(2024)}]{Larsson2024}%
  \BibitemOpen
  \bibfield  {author} {\bibinfo {author} {\bibfnamefont {H.~R.}\ \bibnamefont
  {Larsson}},\ }\bibfield  {title} {\bibinfo {title} {A tensor network view of
  multilayer multiconfiguration time-dependent hartree methods},\ }\href
  {https://doi.org/10.1080/00268976.2024.2306881} {\bibfield  {journal}
  {\bibinfo  {journal} {Molecular Physics}\ }\textbf {\bibinfo {volume} {0}},\
  \bibinfo {pages} {e2306881} (\bibinfo {year} {2024})},\ \Eprint
  {https://arxiv.org/abs/https://doi.org/10.1080/00268976.2024.2306881}
  {https://doi.org/10.1080/00268976.2024.2306881} \BibitemShut {NoStop}%
\bibitem [{\citenamefont {Baxter}(1968)}]{Baxter1968}%
  \BibitemOpen
  \bibfield  {author} {\bibinfo {author} {\bibfnamefont {R.~J.}\ \bibnamefont
  {Baxter}},\ }\bibfield  {title} {\bibinfo {title} {Dimers on a rectangular
  lattice},\ }\href {https://doi.org/10.1063/1.1664623} {\bibfield  {journal}
  {\bibinfo  {journal} {J. Math. Phys}\ }\textbf {\bibinfo {volume} {9}},\
  \bibinfo {pages} {650} (\bibinfo {year} {1968})}\BibitemShut {NoStop}%
\bibitem [{\citenamefont {{\"O}stlund}\ and\ \citenamefont
  {Rommer}(1995)}]{OstlundR1995}%
  \BibitemOpen
  \bibfield  {author} {\bibinfo {author} {\bibfnamefont {S.}~\bibnamefont
  {{\"O}stlund}}\ and\ \bibinfo {author} {\bibfnamefont {S.}~\bibnamefont
  {Rommer}},\ }\bibfield  {title} {\bibinfo {title} {Thermodynamic limit of
  density matrix renormalization},\ }\href
  {https://doi.org/10.1103/PhysRevLett.75.3537} {\bibfield  {journal} {\bibinfo
   {journal} {Phys. Rev. Lett.}\ }\textbf {\bibinfo {volume} {75}},\ \bibinfo
  {pages} {3537} (\bibinfo {year} {1995})}\BibitemShut {NoStop}%
\bibitem [{\citenamefont {Rommer}\ and\ \citenamefont
  {{\"O}stlund}(1997)}]{RommerO1997}%
  \BibitemOpen
  \bibfield  {author} {\bibinfo {author} {\bibfnamefont {S.}~\bibnamefont
  {Rommer}}\ and\ \bibinfo {author} {\bibfnamefont {S.}~\bibnamefont
  {{\"O}stlund}},\ }\bibfield  {title} {\bibinfo {title} {Class of ansatz wave
  functions for one-dimensional spin systems and their relation to the density
  matrix renormalization group},\ }\href
  {https://doi.org/10.1103/PhysRevB.55.2164} {\bibfield  {journal} {\bibinfo
  {journal} {Phys. Rev. B}\ }\textbf {\bibinfo {volume} {55}},\ \bibinfo
  {pages} {2164} (\bibinfo {year} {1997})}\BibitemShut {NoStop}%
\bibitem [{\citenamefont {Nishino}\ \emph {et~al.}(2001)\citenamefont
  {Nishino}, \citenamefont {Hieida}, \citenamefont {Okunishi}, \citenamefont
  {Maeshima}, \citenamefont {Akutsu},\ and\ \citenamefont
  {Gendiar}}]{NishinoHOMAG2001}%
  \BibitemOpen
  \bibfield  {author} {\bibinfo {author} {\bibfnamefont {T.}~\bibnamefont
  {Nishino}}, \bibinfo {author} {\bibfnamefont {Y.}~\bibnamefont {Hieida}},
  \bibinfo {author} {\bibfnamefont {K.}~\bibnamefont {Okunishi}}, \bibinfo
  {author} {\bibfnamefont {N.}~\bibnamefont {Maeshima}}, \bibinfo {author}
  {\bibfnamefont {Y.}~\bibnamefont {Akutsu}},\ and\ \bibinfo {author}
  {\bibfnamefont {A.}~\bibnamefont {Gendiar}},\ }\bibfield  {title} {\bibinfo
  {title} {{Two-Dimensional Tensor Product Variational Formulation}},\ }\href
  {https://doi.org/10.1143/PTP.105.409} {\bibfield  {journal} {\bibinfo
  {journal} {Prog. Theor. Phys.}\ }\textbf {\bibinfo {volume} {105}},\ \bibinfo
  {pages} {409} (\bibinfo {year} {2001})}\BibitemShut {NoStop}%
\bibitem [{\citenamefont {Gendiar}\ \emph {et~al.}(2003)\citenamefont
  {Gendiar}, \citenamefont {Maeshima},\ and\ \citenamefont
  {Nishino}}]{GendNishino2003}%
  \BibitemOpen
  \bibfield  {author} {\bibinfo {author} {\bibfnamefont {A.}~\bibnamefont
  {Gendiar}}, \bibinfo {author} {\bibfnamefont {N.}~\bibnamefont {Maeshima}},\
  and\ \bibinfo {author} {\bibfnamefont {T.}~\bibnamefont {Nishino}},\
  }\bibfield  {title} {\bibinfo {title} {{Stable Optimization of a Tensor
  Product Variational State}},\ }\href {https://doi.org/10.1143/PTP.110.691}
  {\bibfield  {journal} {\bibinfo  {journal} {Prog. Theor. Phys.}\ }\textbf
  {\bibinfo {volume} {110}},\ \bibinfo {pages} {691} (\bibinfo {year}
  {2003})}\BibitemShut {NoStop}%
\bibitem [{\citenamefont {Verstraete}\ and\ \citenamefont
  {Cirac}(2004)}]{VerstraeteC2004}%
  \BibitemOpen
  \bibfield  {author} {\bibinfo {author} {\bibfnamefont {F.}~\bibnamefont
  {Verstraete}}\ and\ \bibinfo {author} {\bibfnamefont {J.~I.}\ \bibnamefont
  {Cirac}},\ }\bibfield  {title} {\bibinfo {title} {Valence-bond states for
  quantum computation},\ }\href {https://doi.org/10.1103/PhysRevA.70.060302}
  {\bibfield  {journal} {\bibinfo  {journal} {Phys. Rev. A}\ }\textbf {\bibinfo
  {volume} {70}},\ \bibinfo {pages} {060302} (\bibinfo {year}
  {2004})}\BibitemShut {NoStop}%
\bibitem [{\citenamefont {Verstraete}\ \emph {et~al.}(2006)\citenamefont
  {Verstraete}, \citenamefont {Wolf}, \citenamefont {P{\'e}rez-Garc{\'{\i}}a},\
  and\ \citenamefont {Cirac}}]{VerstraeteWPC2006}%
  \BibitemOpen
  \bibfield  {author} {\bibinfo {author} {\bibfnamefont {F.}~\bibnamefont
  {Verstraete}}, \bibinfo {author} {\bibfnamefont {M.~M.}\ \bibnamefont
  {Wolf}}, \bibinfo {author} {\bibfnamefont {D.}~\bibnamefont
  {P{\'e}rez-Garc{\'{\i}}a}},\ and\ \bibinfo {author} {\bibfnamefont {J.~I.}\
  \bibnamefont {Cirac}},\ }\bibfield  {title} {\bibinfo {title} {Criticality,
  the area law, and the computational power of projected entangled pair
  states},\ }\href {https://doi.org/10.1103/PhysRevLett.96.220601} {\bibfield
  {journal} {\bibinfo  {journal} {Phys. Rev. Lett.}\ }\textbf {\bibinfo
  {volume} {96}},\ \bibinfo {pages} {220601} (\bibinfo {year}
  {2006})}\BibitemShut {NoStop}%
\bibitem [{\citenamefont {Vidal}(2007)}]{Vidal2007}%
  \BibitemOpen
  \bibfield  {author} {\bibinfo {author} {\bibfnamefont {G.}~\bibnamefont
  {Vidal}},\ }\bibfield  {title} {\bibinfo {title} {Entanglement
  renormalization},\ }\href {https://doi.org/10.1103/PhysRevLett.99.220405}
  {\bibfield  {journal} {\bibinfo  {journal} {Phys. Rev. Lett.}\ }\textbf
  {\bibinfo {volume} {99}},\ \bibinfo {pages} {220405} (\bibinfo {year}
  {2007})}\BibitemShut {NoStop}%
\bibitem [{\citenamefont {Evenbly}\ and\ \citenamefont
  {Vidal}(2009)}]{EvenblyV2009}%
  \BibitemOpen
  \bibfield  {author} {\bibinfo {author} {\bibfnamefont {G.}~\bibnamefont
  {Evenbly}}\ and\ \bibinfo {author} {\bibfnamefont {G.}~\bibnamefont
  {Vidal}},\ }\bibfield  {title} {\bibinfo {title} {Algorithms for entanglement
  renormalization},\ }\href {https://doi.org/10.1103/PhysRevB.79.144108}
  {\bibfield  {journal} {\bibinfo  {journal} {Phys. Rev. B}\ }\textbf {\bibinfo
  {volume} {79}},\ \bibinfo {pages} {144108} (\bibinfo {year}
  {2009})}\BibitemShut {NoStop}%
\bibitem [{\citenamefont {Seki}\ \emph {et~al.}(2021)\citenamefont {Seki},
  \citenamefont {Hikihara},\ and\ \citenamefont {Okunishi}}]{SekiHO2021}%
  \BibitemOpen
  \bibfield  {author} {\bibinfo {author} {\bibfnamefont {K.}~\bibnamefont
  {Seki}}, \bibinfo {author} {\bibfnamefont {T.}~\bibnamefont {Hikihara}},\
  and\ \bibinfo {author} {\bibfnamefont {K.}~\bibnamefont {Okunishi}},\
  }\bibfield  {title} {\bibinfo {title} {Entanglement-based tensor-network
  strong-disorder renormalization group},\ }\href
  {https://doi.org/10.1103/PhysRevB.104.134405} {\bibfield  {journal} {\bibinfo
   {journal} {Phys. Rev. B}\ }\textbf {\bibinfo {volume} {104}},\ \bibinfo
  {pages} {134405} (\bibinfo {year} {2021})}\BibitemShut {NoStop}%
\bibitem [{\citenamefont {Hikihara}\ \emph {et~al.}(2023)\citenamefont
  {Hikihara}, \citenamefont {Ueda}, \citenamefont {Okunishi}, \citenamefont
  {Harada},\ and\ \citenamefont {Nishino}}]{HikiharaUOHN2023a}%
  \BibitemOpen
  \bibfield  {author} {\bibinfo {author} {\bibfnamefont {T.}~\bibnamefont
  {Hikihara}}, \bibinfo {author} {\bibfnamefont {H.}~\bibnamefont {Ueda}},
  \bibinfo {author} {\bibfnamefont {K.}~\bibnamefont {Okunishi}}, \bibinfo
  {author} {\bibfnamefont {K.}~\bibnamefont {Harada}},\ and\ \bibinfo {author}
  {\bibfnamefont {T.}~\bibnamefont {Nishino}},\ }\bibfield  {title} {\bibinfo
  {title} {Automatic structural optimization of tree tensor networks},\ }\href
  {https://doi.org/10.1103/PhysRevResearch.5.013031} {\bibfield  {journal}
  {\bibinfo  {journal} {Phys. Rev. Res.}\ }\textbf {\bibinfo {volume} {5}},\
  \bibinfo {pages} {013031} (\bibinfo {year} {2023})}\BibitemShut {NoStop}%
\bibitem [{\citenamefont {Okunishi}\ \emph {et~al.}(2023)\citenamefont
  {Okunishi}, \citenamefont {Ueda},\ and\ \citenamefont
  {Nishino}}]{OkunishiUN2023}%
  \BibitemOpen
  \bibfield  {author} {\bibinfo {author} {\bibfnamefont {K.}~\bibnamefont
  {Okunishi}}, \bibinfo {author} {\bibfnamefont {H.}~\bibnamefont {Ueda}},\
  and\ \bibinfo {author} {\bibfnamefont {T.}~\bibnamefont {Nishino}},\
  }\bibfield  {title} {\bibinfo {title} {{Entanglement bipartitioning and tree
  tensor networks}},\ }\href {https://doi.org/10.1093/ptep/ptad018} {\bibfield
  {journal} {\bibinfo  {journal} {Progress of Theoretical and Experimental
  Physics}\ }\textbf {\bibinfo {volume} {2023}},\ \bibinfo {pages} {023A02}
  (\bibinfo {year} {2023})},\ \Eprint
  {https://arxiv.org/abs/https://academic.oup.com/ptep/article-pdf/2023/2/023A02/49294184/ptad018.pdf}
  {https://academic.oup.com/ptep/article-pdf/2023/2/023A02/49294184/ptad018.pdf}
  \BibitemShut {NoStop}%
\bibitem [{\citenamefont {Chan}\ and\ \citenamefont
  {Head-Gordon}(2002)}]{ChanHG2002}%
  \BibitemOpen
  \bibfield  {author} {\bibinfo {author} {\bibfnamefont {G.~K.-L.}\
  \bibnamefont {Chan}}\ and\ \bibinfo {author} {\bibfnamefont {M.}~\bibnamefont
  {Head-Gordon}},\ }\bibfield  {title} {\bibinfo {title} {{Highly correlated
  calculations with a polynomial cost algorithm: A study of the density matrix
  renormalization group}},\ }\href {https://doi.org/10.1063/1.1449459}
  {\bibfield  {journal} {\bibinfo  {journal} {The Journal of Chemical Physics}\
  }\textbf {\bibinfo {volume} {116}},\ \bibinfo {pages} {4462} (\bibinfo {year}
  {2002})},\ \Eprint
  {https://arxiv.org/abs/https://pubs.aip.org/aip/jcp/article-pdf/116/11/4462/19222618/4462\_1\_online.pdf}
  {https://pubs.aip.org/aip/jcp/article-pdf/116/11/4462/19222618/4462\_1\_online.pdf}
  \BibitemShut {NoStop}%
\bibitem [{\citenamefont {Legeza}\ and\ \citenamefont
  {S\'olyom}(2003)}]{LegezaS2003}%
  \BibitemOpen
  \bibfield  {author} {\bibinfo {author} {\bibfnamefont {O.}~\bibnamefont
  {Legeza}}\ and\ \bibinfo {author} {\bibfnamefont {J.}~\bibnamefont
  {S\'olyom}},\ }\bibfield  {title} {\bibinfo {title} {Optimizing the
  density-matrix renormalization group method using quantum information
  entropy},\ }\href {https://doi.org/10.1103/PhysRevB.68.195116} {\bibfield
  {journal} {\bibinfo  {journal} {Phys. Rev. B}\ }\textbf {\bibinfo {volume}
  {68}},\ \bibinfo {pages} {195116} (\bibinfo {year} {2003})}\BibitemShut
  {NoStop}%
\bibitem [{\citenamefont {Moritz}\ \emph {et~al.}(2004)\citenamefont {Moritz},
  \citenamefont {Hess},\ and\ \citenamefont {Reiher}}]{MoritzHR2004}%
  \BibitemOpen
  \bibfield  {author} {\bibinfo {author} {\bibfnamefont {G.}~\bibnamefont
  {Moritz}}, \bibinfo {author} {\bibfnamefont {B.~A.}\ \bibnamefont {Hess}},\
  and\ \bibinfo {author} {\bibfnamefont {M.}~\bibnamefont {Reiher}},\
  }\bibfield  {title} {\bibinfo {title} {{Convergence behavior of the
  density-matrix renormalization group algorithm for optimized orbital
  orderings}},\ }\href {https://doi.org/10.1063/1.1824891} {\bibfield
  {journal} {\bibinfo  {journal} {The Journal of Chemical Physics}\ }\textbf
  {\bibinfo {volume} {122}},\ \bibinfo {pages} {024107} (\bibinfo {year}
  {2004})},\ \Eprint
  {https://arxiv.org/abs/https://pubs.aip.org/aip/jcp/article-pdf/doi/10.1063/1.1824891/10872404/024107\_1\_online.pdf}
  {https://pubs.aip.org/aip/jcp/article-pdf/doi/10.1063/1.1824891/10872404/024107\_1\_online.pdf}
  \BibitemShut {NoStop}%
\bibitem [{\citenamefont {Legeza}\ \emph {et~al.}(2015)\citenamefont {Legeza},
  \citenamefont {Veis}, \citenamefont {Poves},\ and\ \citenamefont
  {Dukelsky}}]{LegezaVPD2015}%
  \BibitemOpen
  \bibfield  {author} {\bibinfo {author} {\bibfnamefont {O.}~\bibnamefont
  {Legeza}}, \bibinfo {author} {\bibfnamefont {L.}~\bibnamefont {Veis}},
  \bibinfo {author} {\bibfnamefont {A.}~\bibnamefont {Poves}},\ and\ \bibinfo
  {author} {\bibfnamefont {J.}~\bibnamefont {Dukelsky}},\ }\bibfield  {title}
  {\bibinfo {title} {Advanced density matrix renormalization group method for
  nuclear structure calculations},\ }\href
  {https://doi.org/10.1103/PhysRevC.92.051303} {\bibfield  {journal} {\bibinfo
  {journal} {Phys. Rev. C}\ }\textbf {\bibinfo {volume} {92}},\ \bibinfo
  {pages} {051303} (\bibinfo {year} {2015})}\BibitemShut {NoStop}%
\bibitem [{\citenamefont {Li}\ \emph {et~al.}(2022)\citenamefont {Li},
  \citenamefont {Ren}, \citenamefont {Yang},\ and\ \citenamefont
  {Shuai}}]{LiRYS2022}%
  \BibitemOpen
  \bibfield  {author} {\bibinfo {author} {\bibfnamefont {W.}~\bibnamefont
  {Li}}, \bibinfo {author} {\bibfnamefont {J.}~\bibnamefont {Ren}}, \bibinfo
  {author} {\bibfnamefont {H.}~\bibnamefont {Yang}},\ and\ \bibinfo {author}
  {\bibfnamefont {Z.}~\bibnamefont {Shuai}},\ }\bibfield  {title} {\bibinfo
  {title} {On the fly swapping algorithm for ordering of degrees of freedom in
  density matrix renormalization group},\ }\href
  {https://doi.org/10.1088/1361-648X/ac640e} {\bibfield  {journal} {\bibinfo
  {journal} {Journal of Physics: Condensed Matter}\ }\textbf {\bibinfo {volume}
  {34}},\ \bibinfo {pages} {254003} (\bibinfo {year} {2022})}\BibitemShut
  {NoStop}%
\bibitem [{\citenamefont {Larsson}(2019)}]{Larsson2019}%
  \BibitemOpen
  \bibfield  {author} {\bibinfo {author} {\bibfnamefont {H.~R.}\ \bibnamefont
  {Larsson}},\ }\bibfield  {title} {\bibinfo {title} {Computing vibrational
  eigenstates with tree tensor network states (ttns)},\ }\href
  {https://doi.org/10.1063/1.5130390} {\bibfield  {journal} {\bibinfo
  {journal} {The Journal of Chemical Physics}\ }\textbf {\bibinfo {volume}
  {151}},\ \bibinfo {pages} {204102} (\bibinfo {year} {2019})},\ \Eprint
  {https://arxiv.org/abs/https://doi.org/10.1063/1.5130390}
  {https://doi.org/10.1063/1.5130390} \BibitemShut {NoStop}%
\bibitem [{\citenamefont {Ma}\ \emph {et~al.}(1979)\citenamefont {Ma},
  \citenamefont {Dasgupta},\ and\ \citenamefont {Hu}}]{MaDH1979}%
  \BibitemOpen
  \bibfield  {author} {\bibinfo {author} {\bibfnamefont {S.-k.}\ \bibnamefont
  {Ma}}, \bibinfo {author} {\bibfnamefont {C.}~\bibnamefont {Dasgupta}},\ and\
  \bibinfo {author} {\bibfnamefont {C.-k.}\ \bibnamefont {Hu}},\ }\bibfield
  {title} {\bibinfo {title} {Random antiferromagnetic chain},\ }\href
  {https://doi.org/10.1103/PhysRevLett.43.1434} {\bibfield  {journal} {\bibinfo
   {journal} {Phys. Rev. Lett.}\ }\textbf {\bibinfo {volume} {43}},\ \bibinfo
  {pages} {1434} (\bibinfo {year} {1979})}\BibitemShut {NoStop}%
\bibitem [{\citenamefont {Dasgupta}\ and\ \citenamefont
  {Ma}(1980)}]{DasguptaM1980}%
  \BibitemOpen
  \bibfield  {author} {\bibinfo {author} {\bibfnamefont {C.}~\bibnamefont
  {Dasgupta}}\ and\ \bibinfo {author} {\bibfnamefont {S.-k.}\ \bibnamefont
  {Ma}},\ }\bibfield  {title} {\bibinfo {title} {Low-temperature properties of
  the random heisenberg antiferromagnetic chain},\ }\href
  {https://doi.org/10.1103/PhysRevB.22.1305} {\bibfield  {journal} {\bibinfo
  {journal} {Phys. Rev. B}\ }\textbf {\bibinfo {volume} {22}},\ \bibinfo
  {pages} {1305} (\bibinfo {year} {1980})}\BibitemShut {NoStop}%
\bibitem [{\citenamefont {Hikihara}\ \emph {et~al.}(1999)\citenamefont
  {Hikihara}, \citenamefont {Furusaki},\ and\ \citenamefont
  {Sigrist}}]{HikiharaFS1999}%
  \BibitemOpen
  \bibfield  {author} {\bibinfo {author} {\bibfnamefont {T.}~\bibnamefont
  {Hikihara}}, \bibinfo {author} {\bibfnamefont {A.}~\bibnamefont {Furusaki}},\
  and\ \bibinfo {author} {\bibfnamefont {M.}~\bibnamefont {Sigrist}},\
  }\bibfield  {title} {\bibinfo {title} {Numerical renormalization-group study
  of spin correlations in one-dimensional random spin chains},\ }\href
  {https://doi.org/10.1103/PhysRevB.60.12116} {\bibfield  {journal} {\bibinfo
  {journal} {Phys. Rev. B}\ }\textbf {\bibinfo {volume} {60}},\ \bibinfo
  {pages} {12116} (\bibinfo {year} {1999})}\BibitemShut {NoStop}%
\bibitem [{\citenamefont {Goldsborough}\ and\ \citenamefont
  {R{\"o}mer}(2014)}]{GoldsboroughR2014}%
  \BibitemOpen
  \bibfield  {author} {\bibinfo {author} {\bibfnamefont {A.~M.}\ \bibnamefont
  {Goldsborough}}\ and\ \bibinfo {author} {\bibfnamefont {R.~A.}\ \bibnamefont
  {R{\"o}mer}},\ }\bibfield  {title} {\bibinfo {title} {Self-assembling tensor
  networks and holography in disordered spin chains},\ }\href
  {https://doi.org/10.1103/PhysRevB.89.214203} {\bibfield  {journal} {\bibinfo
  {journal} {Phys. Rev. B}\ }\textbf {\bibinfo {volume} {89}},\ \bibinfo
  {pages} {214203} (\bibinfo {year} {2014})}\BibitemShut {NoStop}%
\bibitem [{\citenamefont {Lin}\ \emph {et~al.}(2017)\citenamefont {Lin},
  \citenamefont {Kao}, \citenamefont {Chen},\ and\ \citenamefont
  {Lin}}]{LinKao_2017}%
  \BibitemOpen
  \bibfield  {author} {\bibinfo {author} {\bibfnamefont {Y.-P.}\ \bibnamefont
  {Lin}}, \bibinfo {author} {\bibfnamefont {Y.-J.}\ \bibnamefont {Kao}},
  \bibinfo {author} {\bibfnamefont {P.}~\bibnamefont {Chen}},\ and\ \bibinfo
  {author} {\bibfnamefont {Y.-C.}\ \bibnamefont {Lin}},\ }\bibfield  {title}
  {\bibinfo {title} {Griffiths singularities in the random quantum ising
  antiferromagnet: A tree tensor network renormalization group study},\ }\href
  {https://doi.org/10.1103/PhysRevB.96.064427} {\bibfield  {journal} {\bibinfo
  {journal} {Phys. Rev. B}\ }\textbf {\bibinfo {volume} {96}},\ \bibinfo
  {pages} {064427} (\bibinfo {year} {2017})}\BibitemShut {NoStop}%
\bibitem [{\citenamefont {Seki}\ \emph {et~al.}(2020)\citenamefont {Seki},
  \citenamefont {Hikihara},\ and\ \citenamefont {Okunishi}}]{SekiHO2020}%
  \BibitemOpen
  \bibfield  {author} {\bibinfo {author} {\bibfnamefont {K.}~\bibnamefont
  {Seki}}, \bibinfo {author} {\bibfnamefont {T.}~\bibnamefont {Hikihara}},\
  and\ \bibinfo {author} {\bibfnamefont {K.}~\bibnamefont {Okunishi}},\
  }\bibfield  {title} {\bibinfo {title} {Tensor-network strong-disorder
  renormalization groups for random quantum spin systems in two dimensions},\
  }\href {https://doi.org/10.1103/PhysRevB.102.144439} {\bibfield  {journal}
  {\bibinfo  {journal} {Phys. Rev. B}\ }\textbf {\bibinfo {volume} {102}},\
  \bibinfo {pages} {144439} (\bibinfo {year} {2020})}\BibitemShut {NoStop}%
\bibitem [{\citenamefont {Hikihara}\ \emph {et~al.}(2024)\citenamefont
  {Hikihara}, \citenamefont {Ueda}, \citenamefont {Okunishi}, \citenamefont
  {Harada},\ and\ \citenamefont {Nishino}}]{HikiharaUOHN2023b}%
  \BibitemOpen
  \bibfield  {author} {\bibinfo {author} {\bibfnamefont {T.}~\bibnamefont
  {Hikihara}}, \bibinfo {author} {\bibfnamefont {H.}~\bibnamefont {Ueda}},
  \bibinfo {author} {\bibfnamefont {K.}~\bibnamefont {Okunishi}}, \bibinfo
  {author} {\bibfnamefont {K.}~\bibnamefont {Harada}},\ and\ \bibinfo {author}
  {\bibfnamefont {T.}~\bibnamefont {Nishino}},\ }\href
  {https://arxiv.org/abs/2401.16000} {\bibinfo {title} {Visualization of
  entanglement geometry by structural optimization of tree tensor network}}
  (\bibinfo {year} {2024}),\ \Eprint {https://arxiv.org/abs/2401.16000}
  {arXiv:2401.16000 [cond-mat.stat-mech]} \BibitemShut {NoStop}%
\bibitem [{\citenamefont {Shi}\ \emph {et~al.}(2006)\citenamefont {Shi},
  \citenamefont {Duan},\ and\ \citenamefont {Vidal}}]{ShiDV2006}%
  \BibitemOpen
  \bibfield  {author} {\bibinfo {author} {\bibfnamefont {Y.-Y.}\ \bibnamefont
  {Shi}}, \bibinfo {author} {\bibfnamefont {L.-M.}\ \bibnamefont {Duan}},\ and\
  \bibinfo {author} {\bibfnamefont {G.}~\bibnamefont {Vidal}},\ }\bibfield
  {title} {\bibinfo {title} {Classical simulation of quantum many-body systems
  with a tree tensor network},\ }\href
  {https://doi.org/10.1103/PhysRevA.74.022320} {\bibfield  {journal} {\bibinfo
  {journal} {Phys. Rev. A}\ }\textbf {\bibinfo {volume} {74}},\ \bibinfo
  {pages} {022320} (\bibinfo {year} {2006})}\BibitemShut {NoStop}%
\bibitem [{Not()}]{Note_isometry_update}%
  \BibitemOpen
  \bibinfo {note} {In the actual computation, one of the two isometries will be
  included in the central area in the next step and does not necessarily need
  to be updated.}\BibitemShut {Stop}%
\bibitem [{\citenamefont {{von Delft}}\ and\ \citenamefont
  {Ralph}(2001)}]{DelftR2001}%
  \BibitemOpen
  \bibfield  {author} {\bibinfo {author} {\bibfnamefont {J.}~\bibnamefont {{von
  Delft}}}\ and\ \bibinfo {author} {\bibfnamefont {D.}~\bibnamefont {Ralph}},\
  }\bibfield  {title} {\bibinfo {title} {Spectroscopy of discrete energy levels
  in ultrasmall metallic grains},\ }\href
  {https://doi.org/https://doi.org/10.1016/S0370-1573(00)00099-5} {\bibfield
  {journal} {\bibinfo  {journal} {Physics Reports}\ }\textbf {\bibinfo {volume}
  {345}},\ \bibinfo {pages} {61} (\bibinfo {year} {2001})}\BibitemShut
  {NoStop}%
\bibitem [{\citenamefont {Asorey}\ \emph {et~al.}(2002)\citenamefont {Asorey},
  \citenamefont {Falceto},\ and\ \citenamefont {Sierra}}]{AsoreyFS2002}%
  \BibitemOpen
  \bibfield  {author} {\bibinfo {author} {\bibfnamefont {M.}~\bibnamefont
  {Asorey}}, \bibinfo {author} {\bibfnamefont {F.}~\bibnamefont {Falceto}},\
  and\ \bibinfo {author} {\bibfnamefont {G.}~\bibnamefont {Sierra}},\
  }\bibfield  {title} {\bibinfo {title} {Chern-simons theory and bcs
  superconductivity},\ }\href
  {https://doi.org/https://doi.org/10.1016/S0550-3213(01)00614-9} {\bibfield
  {journal} {\bibinfo  {journal} {Nuclear Physics B}\ }\textbf {\bibinfo
  {volume} {622}},\ \bibinfo {pages} {593} (\bibinfo {year}
  {2002})}\BibitemShut {NoStop}%
\bibitem [{\citenamefont {Dukelsky}\ \emph {et~al.}(2004)\citenamefont
  {Dukelsky}, \citenamefont {Pittel},\ and\ \citenamefont
  {Sierra}}]{DukelskyPS2004}%
  \BibitemOpen
  \bibfield  {author} {\bibinfo {author} {\bibfnamefont {J.}~\bibnamefont
  {Dukelsky}}, \bibinfo {author} {\bibfnamefont {S.}~\bibnamefont {Pittel}},\
  and\ \bibinfo {author} {\bibfnamefont {G.}~\bibnamefont {Sierra}},\
  }\bibfield  {title} {\bibinfo {title} {Colloquium: Exactly solvable
  richardson-gaudin models for many-body quantum systems},\ }\href
  {https://doi.org/10.1103/RevModPhys.76.643} {\bibfield  {journal} {\bibinfo
  {journal} {Rev. Mod. Phys.}\ }\textbf {\bibinfo {volume} {76}},\ \bibinfo
  {pages} {643} (\bibinfo {year} {2004})}\BibitemShut {NoStop}%
\bibitem [{acc()}]{accuracy_limit}%
  \BibitemOpen
  \bibinfo {note} {The accuracy limit of our calculation is set by the
  conditions that we have used the double precision variables and that we have
  employed the rule in the truncation procedure to discard the bases with
  singular values smaller than $10^{-6}$.}\BibitemShut {Stop}%
\bibitem [{\citenamefont {Hukushima}\ and\ \citenamefont
  {Nemoto}(1996)}]{HukushimaN1996}%
  \BibitemOpen
  \bibfield  {author} {\bibinfo {author} {\bibfnamefont {K.}~\bibnamefont
  {Hukushima}}\ and\ \bibinfo {author} {\bibfnamefont {K.}~\bibnamefont
  {Nemoto}},\ }\bibfield  {title} {\bibinfo {title} {Exchange monte carlo
  method and application to spin glass simulations},\ }\href
  {https://doi.org/10.1143/JPSJ.65.1604} {\bibfield  {journal} {\bibinfo
  {journal} {Journal of the Physical Society of Japan}\ }\textbf {\bibinfo
  {volume} {65}},\ \bibinfo {pages} {1604} (\bibinfo {year} {1996})},\ \Eprint
  {https://arxiv.org/abs/https://doi.org/10.1143/JPSJ.65.1604}
  {https://doi.org/10.1143/JPSJ.65.1604} \BibitemShut {NoStop}%
\bibitem [{\citenamefont {Rozada}\ \emph {et~al.}(2019)\citenamefont {Rozada},
  \citenamefont {Aramon}, \citenamefont {Machta},\ and\ \citenamefont
  {Katzgraber}}]{RozadaAMK2019}%
  \BibitemOpen
  \bibfield  {author} {\bibinfo {author} {\bibfnamefont {I.}~\bibnamefont
  {Rozada}}, \bibinfo {author} {\bibfnamefont {M.}~\bibnamefont {Aramon}},
  \bibinfo {author} {\bibfnamefont {J.}~\bibnamefont {Machta}},\ and\ \bibinfo
  {author} {\bibfnamefont {H.~G.}\ \bibnamefont {Katzgraber}},\ }\bibfield
  {title} {\bibinfo {title} {Effects of setting temperatures in the parallel
  tempering monte carlo algorithm},\ }\href
  {https://doi.org/10.1103/PhysRevE.100.043311} {\bibfield  {journal} {\bibinfo
   {journal} {Phys. Rev. E}\ }\textbf {\bibinfo {volume} {100}},\ \bibinfo
  {pages} {043311} (\bibinfo {year} {2019})}\BibitemShut {NoStop}%
\bibitem [{\citenamefont {Watanabe}\ and\ \citenamefont
  {Ueda}(2024)}]{WatanabeU2024}%
  \BibitemOpen
  \bibfield  {author} {\bibinfo {author} {\bibfnamefont {R.}~\bibnamefont
  {Watanabe}}\ and\ \bibinfo {author} {\bibfnamefont {H.}~\bibnamefont
  {Ueda}},\ }\bibfield  {title} {\bibinfo {title} {Automatic structural search
  of tensor network states including entanglement renormalization},\ }\href
  {https://doi.org/10.1103/PhysRevResearch.6.033259} {\bibfield  {journal}
  {\bibinfo  {journal} {Phys. Rev. Res.}\ }\textbf {\bibinfo {volume} {6}},\
  \bibinfo {pages} {033259} (\bibinfo {year} {2024})}\BibitemShut {NoStop}%
\bibitem [{\citenamefont {Patra}\ \emph {et~al.}(2025)\citenamefont {Patra},
  \citenamefont {Singh},\ and\ \citenamefont {Or\'us}}]{PatraSO2025}%
  \BibitemOpen
  \bibfield  {author} {\bibinfo {author} {\bibfnamefont {S.}~\bibnamefont
  {Patra}}, \bibinfo {author} {\bibfnamefont {S.}~\bibnamefont {Singh}},\ and\
  \bibinfo {author} {\bibfnamefont {R.}~\bibnamefont {Or\'us}},\ }\bibfield
  {title} {\bibinfo {title} {Projected entangled pair states with flexible
  geometry},\ }\href {https://doi.org/10.1103/PhysRevResearch.7.L012002}
  {\bibfield  {journal} {\bibinfo  {journal} {Phys. Rev. Res.}\ }\textbf
  {\bibinfo {volume} {7}},\ \bibinfo {pages} {L012002} (\bibinfo {year}
  {2025})}\BibitemShut {NoStop}%
\bibitem [{\citenamefont {Watanabe}\ \emph {et~al.}(2025)\citenamefont
  {Watanabe}, \citenamefont {Manabe}, \citenamefont {Hikihara},\ and\
  \citenamefont {Ueda}}]{WatanabeMHU2025}%
  \BibitemOpen
  \bibfield  {author} {\bibinfo {author} {\bibfnamefont {R.}~\bibnamefont
  {Watanabe}}, \bibinfo {author} {\bibfnamefont {H.}~\bibnamefont {Manabe}},
  \bibinfo {author} {\bibfnamefont {T.}~\bibnamefont {Hikihara}},\ and\
  \bibinfo {author} {\bibfnamefont {H.}~\bibnamefont {Ueda}},\ }\href
  {https://arxiv.org/abs/2505.05908} {\bibinfo {title} {Ttnopt: Tree tensor
  network package for high-rank tensor compression}} (\bibinfo {year} {2025}),\
  \Eprint {https://arxiv.org/abs/2505.05908} {arXiv:2505.05908 [quant-ph]}
  \BibitemShut {NoStop}%
\bibitem [{\citenamefont {Stoudenmire}\ and\ \citenamefont
  {Schwab}(2016)}]{StoudenmireS2016}%
  \BibitemOpen
  \bibfield  {author} {\bibinfo {author} {\bibfnamefont {E.}~\bibnamefont
  {Stoudenmire}}\ and\ \bibinfo {author} {\bibfnamefont {D.~J.}\ \bibnamefont
  {Schwab}},\ }\bibfield  {title} {\bibinfo {title} {Supervised learning with
  tensor networks},\ }in\ \href
  {https://proceedings.neurips.cc/paper_files/paper/2016/file/5314b9674c86e3f9d1ba25ef9bb32895-Paper.pdf}
  {\emph {\bibinfo {booktitle} {Advances in Neural Information Processing
  Systems}}},\ Vol.~\bibinfo {volume} {29},\ \bibinfo {editor} {edited by\
  \bibinfo {editor} {\bibfnamefont {D.}~\bibnamefont {Lee}}, \bibinfo {editor}
  {\bibfnamefont {M.}~\bibnamefont {Sugiyama}}, \bibinfo {editor}
  {\bibfnamefont {U.}~\bibnamefont {Luxburg}}, \bibinfo {editor} {\bibfnamefont
  {I.}~\bibnamefont {Guyon}},\ and\ \bibinfo {editor} {\bibfnamefont
  {R.}~\bibnamefont {Garnett}}}\ (\bibinfo  {publisher} {Curran Associates,
  Inc.},\ \bibinfo {year} {2016})\BibitemShut {NoStop}%
\bibitem [{\citenamefont {Cheng}\ \emph {et~al.}(2019)\citenamefont {Cheng},
  \citenamefont {Wang}, \citenamefont {Xiang},\ and\ \citenamefont
  {Zhang}}]{ChengWXZ2019}%
  \BibitemOpen
  \bibfield  {author} {\bibinfo {author} {\bibfnamefont {S.}~\bibnamefont
  {Cheng}}, \bibinfo {author} {\bibfnamefont {L.}~\bibnamefont {Wang}},
  \bibinfo {author} {\bibfnamefont {T.}~\bibnamefont {Xiang}},\ and\ \bibinfo
  {author} {\bibfnamefont {P.}~\bibnamefont {Zhang}},\ }\bibfield  {title}
  {\bibinfo {title} {Tree tensor networks for generative modeling},\ }\href
  {https://doi.org/10.1103/PhysRevB.99.155131} {\bibfield  {journal} {\bibinfo
  {journal} {Phys. Rev. B}\ }\textbf {\bibinfo {volume} {99}},\ \bibinfo
  {pages} {155131} (\bibinfo {year} {2019})}\BibitemShut {NoStop}%
\bibitem [{\citenamefont {Shinaoka}\ \emph {et~al.}(2023)\citenamefont
  {Shinaoka}, \citenamefont {Wallerberger}, \citenamefont {Murakami},
  \citenamefont {Nogaki}, \citenamefont {Sakurai}, \citenamefont {Werner},\
  and\ \citenamefont {Kauch}}]{ShinaokaWMNSWK2023}%
  \BibitemOpen
  \bibfield  {author} {\bibinfo {author} {\bibfnamefont {H.}~\bibnamefont
  {Shinaoka}}, \bibinfo {author} {\bibfnamefont {M.}~\bibnamefont
  {Wallerberger}}, \bibinfo {author} {\bibfnamefont {Y.}~\bibnamefont
  {Murakami}}, \bibinfo {author} {\bibfnamefont {K.}~\bibnamefont {Nogaki}},
  \bibinfo {author} {\bibfnamefont {R.}~\bibnamefont {Sakurai}}, \bibinfo
  {author} {\bibfnamefont {P.}~\bibnamefont {Werner}},\ and\ \bibinfo {author}
  {\bibfnamefont {A.}~\bibnamefont {Kauch}},\ }\bibfield  {title} {\bibinfo
  {title} {Multiscale space-time ansatz for correlation functions of quantum
  systems based on quantics tensor trains},\ }\href
  {https://doi.org/10.1103/PhysRevX.13.021015} {\bibfield  {journal} {\bibinfo
  {journal} {Phys. Rev. X}\ }\textbf {\bibinfo {volume} {13}},\ \bibinfo
  {pages} {021015} (\bibinfo {year} {2023})}\BibitemShut {NoStop}%
\bibitem [{\citenamefont {Ritter}\ \emph {et~al.}(2024)\citenamefont {Ritter},
  \citenamefont {N\'u\~nez Fern\'andez}, \citenamefont {Wallerberger},
  \citenamefont {von Delft}, \citenamefont {Shinaoka},\ and\ \citenamefont
  {Waintal}}]{RitterNWDSW2024}%
  \BibitemOpen
  \bibfield  {author} {\bibinfo {author} {\bibfnamefont {M.~K.}\ \bibnamefont
  {Ritter}}, \bibinfo {author} {\bibfnamefont {Y.}~\bibnamefont {N\'u\~nez
  Fern\'andez}}, \bibinfo {author} {\bibfnamefont {M.}~\bibnamefont
  {Wallerberger}}, \bibinfo {author} {\bibfnamefont {J.}~\bibnamefont {von
  Delft}}, \bibinfo {author} {\bibfnamefont {H.}~\bibnamefont {Shinaoka}},\
  and\ \bibinfo {author} {\bibfnamefont {X.}~\bibnamefont {Waintal}},\
  }\bibfield  {title} {\bibinfo {title} {Quantics tensor cross interpolation
  for high-resolution parsimonious representations of multivariate functions},\
  }\href {https://doi.org/10.1103/PhysRevLett.132.056501} {\bibfield  {journal}
  {\bibinfo  {journal} {Phys. Rev. Lett.}\ }\textbf {\bibinfo {volume} {132}},\
  \bibinfo {pages} {056501} (\bibinfo {year} {2024})}\BibitemShut {NoStop}%
\bibitem [{\citenamefont {N\'u\~nez Fern\'andez}\ \emph
  {et~al.}(2024)\citenamefont {N\'u\~nez Fern\'andez}, \citenamefont {Ritter},
  \citenamefont {Jeannin}, \citenamefont {Li}, \citenamefont {Kloss},
  \citenamefont {Louvet}, \citenamefont {Terasaki}, \citenamefont {Parcollet},
  \citenamefont {von Delft}, \citenamefont {Shinaoka},\ and\ \citenamefont
  {Waintal}}]{Fernandez2024}%
  \BibitemOpen
  \bibfield  {author} {\bibinfo {author} {\bibfnamefont {Y.}~\bibnamefont
  {N\'u\~nez Fern\'andez}}, \bibinfo {author} {\bibfnamefont {M.~K.}\
  \bibnamefont {Ritter}}, \bibinfo {author} {\bibfnamefont {M.}~\bibnamefont
  {Jeannin}}, \bibinfo {author} {\bibfnamefont {J.-W.}\ \bibnamefont {Li}},
  \bibinfo {author} {\bibfnamefont {T.}~\bibnamefont {Kloss}}, \bibinfo
  {author} {\bibfnamefont {T.}~\bibnamefont {Louvet}}, \bibinfo {author}
  {\bibfnamefont {S.}~\bibnamefont {Terasaki}}, \bibinfo {author}
  {\bibfnamefont {O.}~\bibnamefont {Parcollet}}, \bibinfo {author}
  {\bibfnamefont {J.}~\bibnamefont {von Delft}}, \bibinfo {author}
  {\bibfnamefont {H.}~\bibnamefont {Shinaoka}},\ and\ \bibinfo {author}
  {\bibfnamefont {X.}~\bibnamefont {Waintal}},\ }\href
  {https://arxiv.org/abs/2407.02454} {\bibinfo {title} {Learning tensor
  networks with tensor cross interpolation: new algorithms and libraries}}
  (\bibinfo {year} {2024}),\ \Eprint {https://arxiv.org/abs/2407.02454}
  {arXiv:2407.02454 [physics.comp-ph]} \BibitemShut {NoStop}%
\bibitem [{\citenamefont {Harada}\ \emph {et~al.}(2025)\citenamefont {Harada},
  \citenamefont {Okubo},\ and\ \citenamefont {Kawashima}}]{Harada2024OK2024}%
  \BibitemOpen
  \bibfield  {author} {\bibinfo {author} {\bibfnamefont {K.}~\bibnamefont
  {Harada}}, \bibinfo {author} {\bibfnamefont {T.}~\bibnamefont {Okubo}},\ and\
  \bibinfo {author} {\bibfnamefont {N.}~\bibnamefont {Kawashima}},\ }\bibfield
  {title} {\bibinfo {title} {Tensor tree learns hidden relational structures in
  data to construct generative models},\ }\href
  {https://doi.org/10.1088/2632-2153/adc2c7} {\bibfield  {journal} {\bibinfo
  {journal} {Machine Learning: Science and Technology}\ }\textbf {\bibinfo
  {volume} {6}},\ \bibinfo {pages} {025002} (\bibinfo {year}
  {2025})}\BibitemShut {NoStop}%
\bibitem [{dat()}]{data_availability}%
  \BibitemOpen
  \bibinfo {note} {Https://github.com/thikihara/TTNdemo.git}\BibitemShut
  {NoStop}%
\bibitem [{\citenamefont {Press}\ \emph {et~al.}()\citenamefont {Press},
  \citenamefont {Teukolsky}, \citenamefont {Vetterling},\ and\ \citenamefont
  {Flannery}}]{NumRecipes}%
  \BibitemOpen
  \bibfield  {author} {\bibinfo {author} {\bibfnamefont {W.~H.}\ \bibnamefont
  {Press}}, \bibinfo {author} {\bibfnamefont {S.~A.}\ \bibnamefont
  {Teukolsky}}, \bibinfo {author} {\bibfnamefont {W.~T.}\ \bibnamefont
  {Vetterling}},\ and\ \bibinfo {author} {\bibfnamefont {B.~P.}\ \bibnamefont
  {Flannery}},\ }\href@noop {} {\bibinfo {title} {{\it Numerical Recipes in
  FORTRAN 77, The Art of Scientific Computing, Second Edition} (cambridge
  university press, new york, 1992)}}\BibitemShut {NoStop}%
\end{thebibliography}

%

\end{document}